\documentclass[aps, floatfix, superscriptaddress]{revtex4}

\usepackage{mathtools}
\usepackage{graphicx}
\usepackage{dcolumn}
\usepackage{bm}
\usepackage{quantikz}

\usepackage{graphicx,amsmath,amssymb,tikz,amsthm} 

\usepackage{mathtools}
\usepackage[linktocpage=true,
  colorlinks=true, 
  pdfborder={0 0 0},
  linkcolor=blue,
  citecolor=red,
  filecolor=yellow,
  urlcolor=blue,
  bookmarks,
  pdfauthor={},
]{hyperref}
\usepackage{orcidlink}

\begin{document}

\title{
Optimized synthesis of circuits for diagonal unitary matrices with reflection symmetry
}

\author{Xinchi Huang\orcidlink{0000-0002-7547-074X}}
\email{kkou@quemix.com}
\affiliation{
Laboratory for Materials and Structures,
Institute of Innovative Research,
Tokyo Institute of Technology,
Yokohama 226-8503,
Japan
}

\affiliation{
Quemix Inc.,
Taiyo Life Nihombashi Building,
2-11-2,
Nihombashi Chuo-ku, 
Tokyo 103-0027, Japan
}

\author{Taichi Kosugi\orcidlink{0000-0003-3379-3361}}
\affiliation{
Laboratory for Materials and Structures,
Institute of Innovative Research,
Tokyo Institute of Technology,
Yokohama 226-8503,
Japan
}

\affiliation{
Quemix Inc.,
Taiyo Life Nihombashi Building,
2-11-2,
Nihombashi Chuo-ku, 
Tokyo 103-0027,
Japan
}

\author{Hirofumi Nishi\orcidlink{0000-0001-5155-6605}}
\affiliation{
Laboratory for Materials and Structures,
Institute of Innovative Research,
Tokyo Institute of Technology,
Yokohama 226-8503,
Japan
}

\affiliation{
Quemix Inc.,
Taiyo Life Nihombashi Building,
2-11-2,
Nihombashi Chuo-ku, 
Tokyo 103-0027,
Japan
}

\author{Yu-ichiro Matsushita\orcidlink{0000-0002-9254-5918}}
\affiliation{
Laboratory for Materials and Structures,
Institute of Innovative Research,
Tokyo Institute of Technology,
Yokohama 226-8503,
Japan
}

\affiliation{
Quemix Inc.,
Taiyo Life Nihombashi Building,
2-11-2,
Nihombashi Chuo-ku, 
Tokyo 103-0027,
Japan
}

\affiliation{Quantum Materials and Applications Research Center,
National Institutes for Quantum Science and Technology (QST),
2-12-1 Ookayama, Meguro-ku, Tokyo 152-8550, Japan
}

\date{\today}

\begin{abstract}
During the noisy intermediate-scale quantum (NISQ) era, it is important to optimize the quantum circuits in circuit depth and gate count, especially entanglement gates, including the CNOT gate. 
Among all the unitary operators, diagonal unitary matrices form a special class that plays a crucial role in many quantum algorithms/subroutines. 
Based on a natural gate set \{CNOT, \textit{R}\textsubscript{\textit{z}}\}, quantum circuits for general diagonal unitary matrices were discussed in several previous works, and an optimal synthesis algorithm was proposed in terms of circuit depth. 
In this paper, we are interested in the implementation of diagonal unitary matrices with reflection symmetry, which has promising applications, including the realization of real-time evolution for first quantized Hamiltonians by quantum circuits. 
Owing to such a symmetric property, we show that the quantum circuit in the existing work can be further simplified and propose a constructive algorithm that optimizes the entanglement gate count. 
Compared to the previous synthesis methods for general diagonal unitary matrices, the quantum circuit by our proposed algorithm achieves nearly half the reduction in both the gate count and circuit depth. 
\end{abstract}

\maketitle 

\section{Introduction}\label{sec:1}

With the rapid development of quantum hardware and devices \cite{Arute.20, Leymann.20, Preskill.18, Wu.21}, the last decade has witnessed an increasing number of publications and improvements in quantum computing and algorithms. 
Quantum circuit synthesis is a process for constructing a quantum circuit that provides an optimized implementation (w.r.t. gate count/depth) for a desired unitary operator in terms of a given set of universal gates, which is a crucial task in quantum computation and information \cite{Shende.06, Vartiainen.04}.
The choice of the set of universal gates may differ, and there could be some constraints on the connectivity of the qubits according to the hardware used \cite{Arute.19, Bartlett.21, Wright.19, Wu.21, Ye.19}.

Among all unitary operators, diagonal unitary matrices form an important class and play a crucial role in many quantum algorithms, such as the oracle query in Grover's algorithm \cite{NC10}, simulation of quantum dynamics, \cite{Abrams.97, Kassal.08, W.14, Zalka98} and decoupling \cite{Nakata.15, Nakata.17}. 
Moreover, the problem of finding an efficient quantum circuit for a given diagonal unitary matrix also serves as a subroutine in the discussion of many decomposition methods, including spectral decomposition, Givens' QR method, etc. \cite{Brennen.06, BM04, Bullock.05, Cybenko01, Shi03, Tucci99}, and has thus been studied over the last decades. 

Based on a natural gate set $\{\mathrm{CNOT}, R_z\}$, $n$-qubit diagonal unitary matrices can be precisely implemented using $2^n$ multifold $z$ rotation gates whose rotation angles are proportional to the Hadamard(-Walsh) transform of the given diagonal matrices \cite{Schuch02, Schuch.03}. 
By sorting Walsh functions in sequency order \cite{Beauchamp84}, 
some CNOT gates can be cancelled, and a quantum circuit of $2^n-2$ CNOT gates with depth $2^{n+1}-3$ was obtained in Refs. \cite{Schuch02, W.14} (see also Ref. \cite{BM04} for the same result with a different theoretical tool). 
As shallow quantum circuits to solve practical problems are eagerly expected during the NISQ era, Refs. \cite{BGK18, ZHL22} improved the quantum circuits in Ref. \cite{W.14} in the sense that the authors achieved nearly half reduction in circuit depth (see Table 1 \cite{ZHL22}) by commuting and parallelizing quantum gates in a collective manner.
Although the number of CNOT gates and circuit depth still exhibit exponential growth for $n$, Ref. \cite{ZHL22} is the best-known result on the exact implementation for general diagonal unitary matrices. 
In contrast, Refs. \cite{Beer.16, W.16} discussed the case of phase-sparse diagonal unitary matrices and showed that such exponential growth could be reduced to $\mathcal{O}(k n^2)$ for diagonal unitaries with $k$ distinct phases. 

In this paper, intrigued by the observation that the asymptotic number of CNOT gates and the circuit depth could be reduced to the number of distinct entries of the diagonal unitary matrices, we aim to implement a kind of symmetric diagonal unitary matrices, without using any ancillary qubits, whose diagonal entries admit reflection symmetry. As applications, we provide two simple numerical results on the first quantized Hamiltonian simulation in one dimension, where we implement the real-time evolution operator for either an even potential (Eckart barrier) or the interactive potential using our proposed circuit. Moreover, we show that the circuit depth and gate count can be nearly halved compared to the result in Ref. \cite{ZHL22}. The circuit depth considered in this study is based on the $\{\mathrm{CNOT}, R_z\}$ implementation. 

The remainder of this paper is organized as follows. 
In Sect.~\ref{sec:2}, we first describe our main purpose precisely and then illustrate the possibility of further reduction from the result \cite{ZHL22} by using some rewriting rules in a simple example of $n=2$ in Sect.~\ref{sec:3}. 
Intrigued by this finding, we propose an automatic CNOT-optimized synthesis algorithm in Sect.~\ref{sec:4}, which is the primary result of this study. 
In Sect.~\ref{sec:appl}, we numerically investigate some Hamiltonian simulations using our proposed algorithm as a subroutine. 
Finally, Sect.~\ref{sec:6} is devoted to the conclusion, in which we summarize the results and provide some concluding remarks. 
\section{Preliminary}
\label{sec:2}

In this section, we describe our problem in detail, which aims to find an efficient quantum circuit to realize a specific type of diagonal unitary matrix with a symmetric property. 

Let $n\in \mathbb{N}:= \{1,2,\ldots\}$, and we intend to efficiently implement the diagonal unitary $2^{2n}\times 2^{2n}$ matrix:  
\begin{equation*}
\mathrm{D}(\bm{\theta}) :=  
\begin{pmatrix}
& e^{\mathrm{i}\theta_{0,0}} & 0 & \cdots & 0 & 0 & 0 & 0 \\
& 0 & e^{\mathrm{i}\theta_{0,1}} & \cdots & 0 & 0 & 0 & 0 \\
& \vdots & \vdots & \ddots & \vdots & \vdots & \vdots & \vdots \\
& 0 & 0 & 0 & e^{\mathrm{i}\theta_{0,2^n-1}} & 0 & 0 & 0 \\
& 0 & 0 & 0 & 0 & e^{\mathrm{i}\theta_{1,0}} & 0 & 0\\
& \vdots & \vdots & \vdots & \vdots & \vdots & \ddots & \vdots\\
& 0 & 0 & 0 & 0 & 0 & 0 & e^{\mathrm{i}\theta_{2^n-1,2^n-1}}
\end{pmatrix},
\end{equation*}
where $\bm{\theta} = (\theta_{0,0},\theta_{0,1}, \ldots, \theta_{0,2^n-1},\theta_{1,0}, \ldots, \theta_{2^n-1,2^n-1})^{\text{T}} \in \mathbb{R}^{2^{2n}}$ satisfies the following symmetric assumption: 
\begin{equation}
\label{intro:eq-theta}
\theta_{j,j^\prime} = \theta_{2^n-1-j,2^n-1-j^\prime}, \quad j,j^\prime = 0,1,\ldots, 2^n-1,
\end{equation}
which indicates that the diagonal entries of $\mathrm{D}(\bm{\theta})$ admit a reflection symmetry. 
If one directly applies the current best method \cite{ZHL22}, one obtains a quantum circuit with $2^{2n}-2$ CNOT gates and a depth of $2^{2n}$. 
In the following context, we deal with this specific type of symmetric diagonal unitary matrix and provide an algorithm to produce the corresponding quantum circuit with $2^{2n-1}+2n-2$ CNOT gates and a depth of $2^{2n-1}+2^{2n-3}$, which achieves nearly half the reduction compared to the previous result for both the number of entanglement gates and the circuit depth. 

\section{Observation}
\label{sec:3}

\begin{figure}
\centering
\resizebox{13.0cm}{!}{
\includegraphics[keepaspectratio]{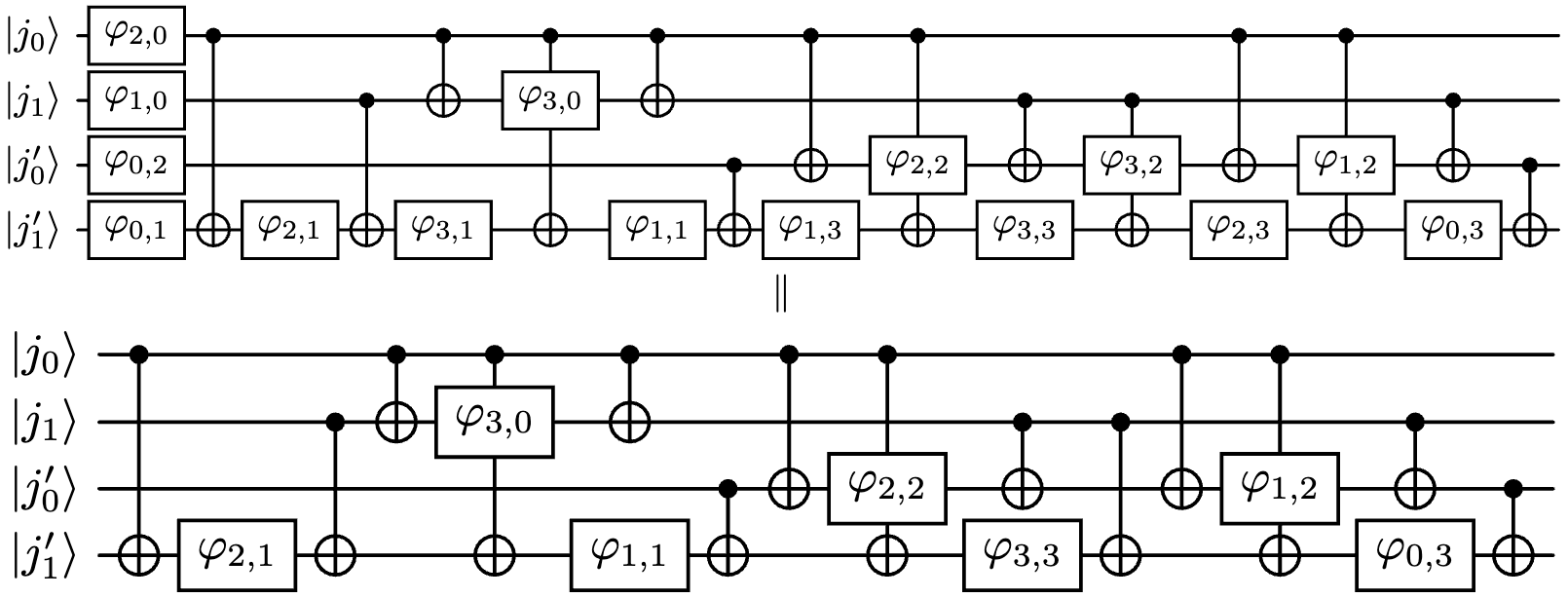}
}%
\caption{Reduction of $R_z$ gates from the circuit in Ref. \cite{ZHL22}. 
Here, $\varphi_{j,j^\prime}$ denotes the rotation angle of the rotation gates. Thanks to Eq. \eqref{intro:eq-theta}, the number of $R_z$ gates is reduced from $15$ to $7$ for $n=2$. }
\label{pre:fig1}
\end{figure}

For an arbitrary $2^{2n}\times 2^{2n}$ diagonal unitary matrix, Zhang et al. \cite{ZHL22} proposed a quantum circuit with $2^{2n}-2$ CNOT gates and $2^{2n}-1$ $R_z$ gates (see the upper circuit in Fig.~\ref{pre:fig1}). 
We start from this existing circuit to demonstrate the possibility of further reduction in both the entanglement gate count and circuit depth under the symmetric assumption \eqref{intro:eq-theta}. 
In fact, regarding Ref. \cite{ZHL22} (or Ref. \cite{W.14} in different notations), we use the matrix of the $2n$-qubit Hadamard transform denoted by $\mathbf{{H}}$ to define a vector $\bm\alpha$:
\begin{equation}
\label{pre:eq-alpha}
\bm{\alpha} \equiv (\alpha_{0,0},\ldots,\alpha_{0,2^n-1},\alpha_{1,0},\ldots,\alpha_{2^n-1,2^n-1})^{\text{T}}
:= \mathbf{
{H}} \bm{\theta}.
\end{equation}
Then the rotation angles of the $R_z$ gates in Fig.~\ref{pre:fig1} are known to be determined by 
$\varphi_{j,j^\prime} = -2\alpha_{j,j^\prime}/\sqrt{2^{2n}}$ for any $j,j^\prime=0,1,\ldots,2^n-1$. 
In the simplest case of $n=1$, the naive observation yields two zero components of $\bm\alpha$: 
\begin{align*}
&\alpha_{0,1} = \left(\theta_{0,0} - \theta_{0,1} + \theta_{1,0} - \theta_{1,1} \right)/2 = 0,\\
&\alpha_{1,0} = \left(\theta_{0,0} + \theta_{0,1} - \theta_{1,0} - \theta_{1,1} \right)/2 = 0,
\end{align*}
where the equalities are derived from the symmetric assumption \eqref{intro:eq-theta}. 
Similarly, for the case of $n=2$, we have $2^{2n-1}=8$ zero components: 
$
\alpha_{0,1}, \alpha_{1,0}, \alpha_{0,2}, \alpha_{2,0}, 
\alpha_{3,1}, \alpha_{1,3}, \alpha_{3,2}, \alpha_{2,3} = 0.
$
This implies that half of $R_z$ gates vanish, and thus, we obtain the circuit equivalence in Fig.~\ref{pre:fig1}. 
We postpone the detailed proof for the case of general $n$ to Appendix \ref{app:subsec-1}. 

\begin{figure}
\centering
\resizebox{13.0cm}{!}{
\includegraphics[keepaspectratio]{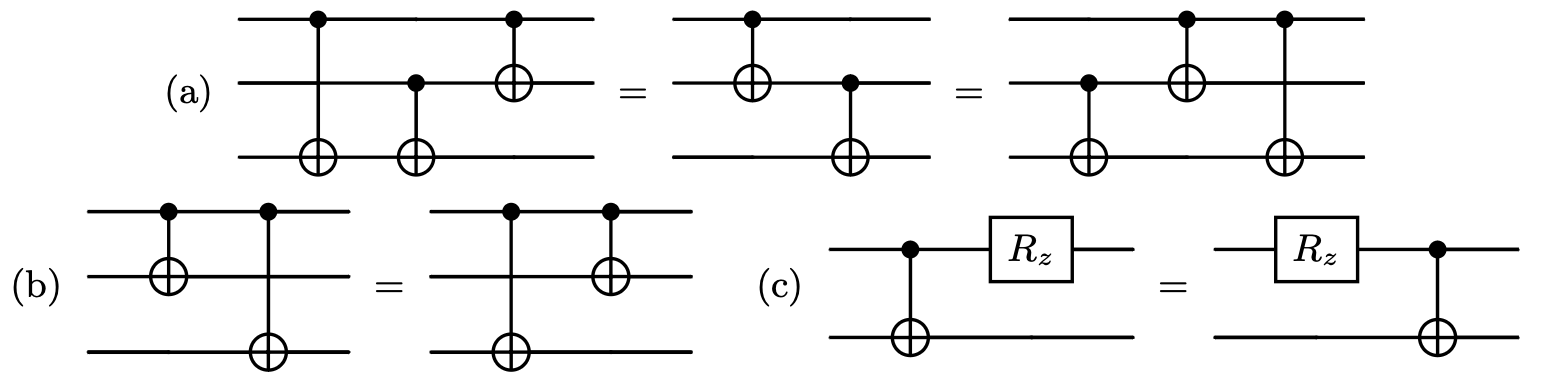}
}%
\caption{Rewriting rules for CNOT gates.}
\label{pre:fig2}
\end{figure}

Next, we consider the reduction of CNOT gates and circuit depth. 
In the example of $n=2$ (i.e.,\! $4$-qubit case), by using some rewriting rules for CNOT gates in Fig.~\ref{pre:fig2}, we manage to further simplify the circuit in Fig.~\ref{pre:fig1} to the one in Fig.~\ref{pre:fig3}. The detailed simplification is provided in Appendix \ref{app:subsec-2}. 
For general $n\ge 2$, one can enlarge the set of the rewriting rules and simplify, by hand calculations or computer software, the circuit in Ref. \cite{ZHL22}, but this is not easy to deal with for large $n$. 
Therefore, we propose an algorithm that automatically achieves such a CNOT-optimized circuit in the following section. 

\begin{figure}
\centering
\resizebox{13.0cm}{!}{
\includegraphics[keepaspectratio]{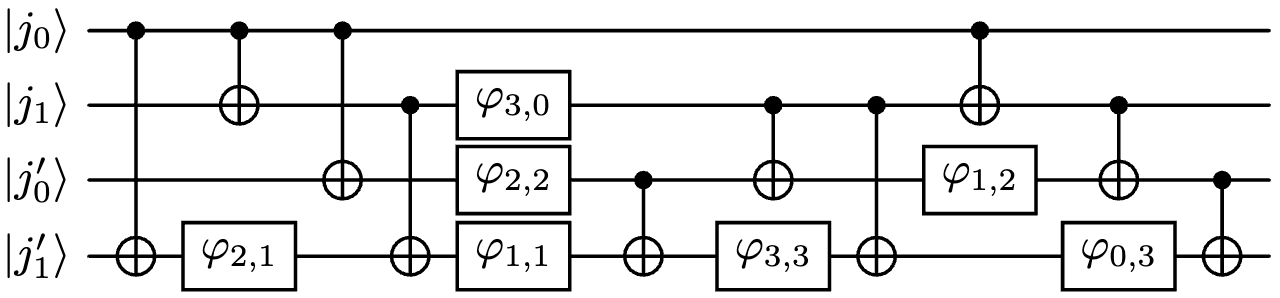}
}%
\caption{Simplified quantum circuit by applying rewriting rules for $n=2$. Note that the third and the fourth CNOT gates can be executed at the same time. Then, the CNOT count is reduced from $14$ to $10$, and the depth is reduced to $10$. }
\label{pre:fig3}
\end{figure}
\section{Method: CNOT-optimized Synthesis Algorithm}
\label{sec:4}

\begin{figure}
\centering
\resizebox{8.0cm}{!}{
\includegraphics[keepaspectratio]{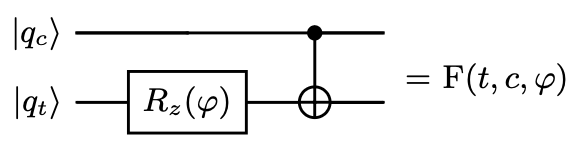}
}%
\caption{An F-block with an $R_z$ gate and a CNOT gate.}
\label{opt:fig1}
\end{figure}

The simplified circuit in Fig.~\ref{pre:fig3} consists of several CNOT gates and some fundamental blocks (pairs of one $R_z$ gate and one CNOT gate). 
Intrigued by the above observation, we provide a constructive algorithm for automatically achieving the optimal number of entanglement gates by building a quantum circuit with the following block in Fig.~\ref{opt:fig1}. 
For the convenience of later discussion, we refer to the fundamental block in Fig.~\ref{opt:fig1} the F-block, and we denote it by $\mathrm{F}(t,c,\varphi)$ with target and control qubits $t,c\in \{0,1,\ldots,2n-1\}$ and a parameter $\varphi\in [0,2\pi]$ describing the rotation angle in the block.  
Moreover, here and henceforth we combine the two registers $\ket{j}_n,\ket{j^\prime}_n$ into $\ket{q}_{2n}$ with the order $\ket{j_m} = \ket{q_m}$, $\ket{j_m^\prime} = \ket{q_{n+m}}$ for $m=0,1,\ldots,n-1$. 

\subsection{Determination of parameters in each block}
\label{sec:4-1}

According to the novel information found in the previous section, the optimal circuit is made up of $2^{2n-1}-1$ F-blocks and $2n-1$ CNOT gates. 
To build the optimal quantum circuit in a constructive manner, we need to determine the parameters in all F-blocks, that is, to create pairs of $R_z$ gates and CNOT gates with suitable target and control qubits. 
To do this, we introduce three useful sequences 
$\{\mathrm{HGC}_m\}_{m=1}^{2n-1}$, $\{\mathrm{CC}_m\}_{m=1}^{2n-1}$ and $\{\mathrm{RAI}_m\}_{m=1}^{2n-1}$. 
The first sequence $\{\mathrm{HGC}_m\}_{m=1}^{2n-1}$ is generated from the Gray code sequence used in Ref. \cite{ZHL22} (see also Refs. \cite{BER76, G53} for an early introduction). Considering that half of the rotation angles are zero, we must modify and use a half Gray code sequence by cutting off the odd (i.e.,\! $(2k-1)$-th) components as follows:
\begin{align*}
&\text{For } m=1: \mathrm{GC}_1 = \{0,1\} \text{, cut off the first component to obtain } \\
&\hspace{1.7cm} \mathrm{HGC}_1 = \{1\}.\\
&\text{For } m\ge 2: \mathrm{GC}_m = \{\mathrm{GC}_{m-1}\{0\},\overline{\mathrm{GC}_{m-1}}\{1\}\} \text{, cut off components} \\ 
&\hspace{1.7cm} \text{of odd indices to obtain } \mathrm{HGC}_m.
\end{align*}
In the last line above, $\mathrm{GC}_{m-1}\{j\}$ is the code by appending $j=0$ or $1$ after the code $\mathrm{GC}_{m-1}$ and $\overline{\mathrm{GC}_{m-1}}$ is the code in the reverse order. 
In this paper, we denote the $m$-bit Gray code and half Gray code by $\mathrm{GC}_{m}, \mathrm{HGC}_{m}$ ($m\ge 1$), respectively, and the half Gray code sequence is used later to construct the sequence $\{\mathrm{RAI}_m\}_{m=1}^{2n-1}$ which determines the rotation angles in F-blocks. 

Next, we introduce the second sequence $\{\mathrm{CC}_m\}_{m=1}^{2n-1}$ (see Eq. (18) in Ref. \cite{ZHL22}), which is employed to determine the control qubits in F-blocks. 
$\{\mathrm{CC}_m\}_{m=1}^{2n-1}$ is defined by iterations: by using $\mathrm{CC}_{m-1}$ we change its last component to $m-1$ and copy the result to derive $\mathrm{CC}_{m}$. That is, 
\begin{align*}
&\text{For } m=1: \mathrm{CC}_1 = \{0\}. \\
&\text{For } m\ge 2: \mathrm{CC}_{\mathrm{temp}}=\mathrm{CC}_{m-1}, \mathrm{CC}_{\mathrm{temp}}[2^{m-2}]=m-1, \\
&\hspace{1.7cm} \mathrm{CC}_m = \{\mathrm{CC}_{\mathrm{temp}},\mathrm{CC}_{\mathrm{temp}}\}. 
\end{align*}
Here, $\mathrm{CC}_{\text{temp}}[j]$ denotes the $(j+1)$-th element. 

Moreover, by using $\{\mathrm{HGC}_m\}_{m=1}^{2n-1}$, we introduce the sequence $\{\mathrm{RAI}_m\}_{m=1}^{2n-1}$, which indicates the indices for the rotation angles. 
In fact, each $\mathrm{RAI}_m$ contains indices in binary representation and is defined by 
$$
\mathrm{RAI}_m := \mathrm{HGC}_m\{1\}\{0^{(2n-1-m)}\},\quad m=1,2,\ldots,2n-1. 
$$
Here, $\{0^{(2n-1-m)}\}$ refers to appending $\{0\}$ for $2n-1-m$ times. 

Now, we are ready to prepare the combination of parameters of the $2^{2n-1}-1$ F-blocks: 
$$
\left(m, \mathrm{CC}_m, \varphi_{\mathrm{RAI}_m}\right), \quad m=1,2,\ldots,2n-1,
$$
where $\varphi_{j_0 \ldots j_{n-1} j_0^\prime \ldots j_{n-1}^\prime} \equiv \varphi_{j,j^\prime}$, that is, we coincide with the index $(j,j^\prime)$ of the rotation angle $\varphi_{j,j^\prime}$ (see Sect.~\ref{sec:3}) with its $2n$-bit binary representation $j_0 \ldots j_{n-1} j_0^\prime \ldots j_{n-1}^\prime$ which is the element in $\{\mathrm{RAI}_m\}_{m=1}^{2n-1}$. 
Here, $\mathrm{CC}_m$ and $\varphi_{\mathrm{RAI}_m}$ are linked by matching the corresponding elements, as listed in Table \ref{tab:3}. Consequently, we succeeded in preparing $2^{2n-1}-1$ F-blocks. 
\begin{table}[htb]
\centering
\caption{Representation of the parameters of the F-blocks in a triangle-like table. }\label{tab:3}
\scalebox{0.6}[0.6]{\tabcolsep=0.1cm 
\begin{tabular}{c|cccccccccccccccc}
\hline
$m$ & \multicolumn{16}{|c}{$(\mathrm{CC}_m, \mathrm{RAI}_m)$} \\
\hline
1 &  &  &  &  &  &  &  &  
&  &  &  &  &  &  &  & (0, 110000)\\
2 &  &  &  &  &  &  &  &  
&  &  &  &  &  &  & (1, 101000) & (1, 011000)\\
3 &  &  &  &  &  &  &  &  
&  &  &  &  & (1, 100100) & (2, 010100) & (1, 111100) & (2, 001100)\\
4 &  &  &  &  &  &  &  &  
& (1, 100010) & (2, 010010) & (1, 111010) & (3, 001010) & (1, 101110) & (2, 011110) & (1, 110110) & (3, 000110)\\
5 & (1, 100001) & (2, 010001) & (1, 111001) & (3, 001001) & (1, 101101) & (2, 011101) & (1, 110101) & (4, 000101) 
& (1, 100111) & (2, 010111) & (1, 111111) & (3, 001111) & (1, 101011) & (2, 011011) & (1, 110011) & (4, 000011)\\
$\cdots$ & $\cdots$ & $\cdots$ & $\cdots$ & $\cdots$ & $\cdots$ & $\cdots$ & $\cdots$ & $\cdots$ 
& $\cdots$ & $\cdots$ & $\cdots$ & $\cdots$ & $\cdots$ & $\cdots$ & $\cdots$ & $\cdots$ \\
\hline
\end{tabular}
}%
\end{table}

Since the F-blocks (as operations) are not commutable in general, we have to suitably determine the order of such blocks. 
Through technical observation, we find that a suitable order can be given according to the triangle-like arrangements in Table \ref{tab:3}. 
Note that in our construction, each F-block corresponds to one entry in the triangle-like table. We arrange the F-blocks in the order of picking up the elements in each column from top to bottom and for all columns from left to right. 
For example, in the case of $n=3$, we have the following order:
{\small
$$
\begin{aligned}
&\mathrm{F}(5,1,\varphi_{100001}), \mathrm{F}(5,2,\varphi_{010001}), 
\mathrm{F}(5,1,\varphi_{111001}), \mathrm{F}(5,3,\varphi_{001001}), \\
&\mathrm{F}(5,1,\varphi_{101101}),\mathrm{F}(5,2,\varphi_{011101}), 
\mathrm{F}(5,1,\varphi_{110101}), \mathrm{F}(5,4,\varphi_{000101}), \\
&\mathrm{F}(4,1,\varphi_{100010}), \mathrm{F}(5,1,\varphi_{100111}), 
\mathrm{F}(4,2,\varphi_{010010}), \mathrm{F}(5,2,\varphi_{010111}), \\
&\mathrm{F}(4,1,\varphi_{111010}), \mathrm{F}(5,1,\varphi_{111111}), 
\mathrm{F}(4,3,\varphi_{001010}),\mathrm{F}(5,3,\varphi_{001111}), \\
&\mathrm{F}(3,1,\varphi_{100100}), \mathrm{F}(4,1,\varphi_{101110}), 
\mathrm{F}(5,1,\varphi_{101011}), \mathrm{F}(3,2,\varphi_{010100}), \\ 
&\mathrm{F}(4,2,\varphi_{011110}), \mathrm{F}(5,2,\varphi_{011011}), 
\mathrm{F}(2,1,\varphi_{101000}), \mathrm{F}(3,1,\varphi_{111100}), \\
&\mathrm{F}(4,1,\varphi_{110110}),\mathrm{F}(5,1,\varphi_{110011}), 
\mathrm{F}(1,0,\varphi_{110000}), \mathrm{F}(2,1,\varphi_{011000}), \\
&\mathrm{F}(3,2,\varphi_{001100}), \mathrm{F}(4,3,\varphi_{000110}), 
\mathrm{F}(5,4,\varphi_{000011}). 
\end{aligned}
$$
}

\subsection{Main algorithm}

We denote the CNOT gate by 
$$
\mathrm{CNOT}(t,c). 
$$
Here, $t,c\in \{0,1,\ldots,2n-1\}$ are the target and control qubits, which means that the CNOT gate is applied to $\ket{q_{t}}$ with the control qubit $\ket{q_c}$. 

We now propose Algorithm 1 as the main contribution of this paper, which directly provides the $\{\mathrm{CNOT}, R_z\}$-based quantum circuit with an optimized CNOT count for the specific symmetric diagonal unitary matrix $\mathrm{D}(\bm{\theta})$ satisfying Eq. \eqref{intro:eq-theta}. 
To clarify the main algorithm, we explain it using the following steps:

\noindent (i) In Line 1, we calculate the rotation angle $\varphi_j$ for each $j$. According to Theorem 1 in Appendix \ref{app:subsec-1}, we only need to calculate the non-zero angles (half of the total angles) whose binary representation $j=[j_0 j_1 \ldots j_{2n-1}]$ satisfies the condition that the sum of all its digits is even. 

\noindent (ii) From Line 2 to Line 10, we generate the necessary sequences $\{\mathrm{CC}_m\}_{m=1}^{2n-1}$, $\{\mathrm{HGC}_m\}_{m=1}^{2n-1}$ and $\{\mathrm{RAI}_m\}_{m=1}^{2n-1}$ as we have illustrated in Sect.~\ref{sec:4-1}. 
Such sequences help us prepare the combination of parameters in the F-blocks. 

\noindent (iii) From Line 11 to Line 30, we construct the desired quantum circuit in several steps. 

\noindent -- Step $0$: We add $2n-1$ CNOT gates to the circuit $\mathrm{QC}$ with the control qubit $\ket{q_0}$ and the target qubits from $\ket{q_{2n-1}}$ to $\ket{q_1}$, that is, $\mathrm{CNOT}(m,0)$ from $m=2n-1$ to $m=1$. 

\noindent -- Step $m$ ($m=1,2,\ldots,2n-2$): We add to the circuit $\mathrm{QC}$ some special block $\mathrm{B}_{2n-1-m}$ (containing $m 2^{2n-2-m}$ F-blocks) where $\mathrm{B}_k$ ($k=1,2,\ldots, 2n-2$) is made up of $\mathrm{F}(k+j,\mathrm{CC}_{k+j}[2^{k+j-1}-2^k+i],\varphi_{\mathrm{RAI}_{k+j}[2^{k+j-1}-2^k+i]})$ for $j=1$ to $j=m$ with $i=1,2,\ldots,2^{k-1}-1,2^{k-1}$. 

\noindent -- Step $2n-1$: We add $2n-1$ F-blocks: 
$\mathrm{F}(j,\mathrm{CC}_{j}[2^{j-1}],\varphi_{\mathrm{RAI}_{j}[2^{j-1}]})$ from $j=1$ to $j=2n-1$ to the circuit $\mathrm{QC}$. 

To avoid confusion, we mention that the proposed quantum circuit in Algorithm 1 provides a realization of $\mathrm{D}(\bm{\theta})$ (see Appendix \ref{app:subsec-3}) up to a global phase. Indeed, if we denote the unitary operator generated from Algorithm 1 by $\mathrm{\tilde D}(\bm{\theta})$, then we have
$$
\mathrm{D}(\bm{\theta}) = e^{-\mathrm{i}\varphi_0/2} \mathrm{\tilde D(\bm{\theta}}), 
$$
where $\varphi_0$ is calculated from $\bm{\theta}$ in Line 1 of Algorithm 1. 
Using this constructive algorithm, we readily find that the number of $R_z$ gates is $2^{2n-1}-1$ and the number of CNOT gates is reduced to $2^{2n-1}-1+2n-1 = 2^{2n-1}+2n-2$, which is nearly half the reduction compared to the circuit in Ref. \cite{ZHL22}. 
\begin{table*}[htb]
\resizebox{15cm}{!}{
\begin{minipage}{\textwidth}
\centering
\begin{tabular}{l}
\hline
\textbf{Algorithm 1: } CNOT-optimized quantum circuit for implementing $\mathrm{D}(\bm{\theta})$ with Eq. \eqref{intro:eq-theta}.\\
\hline
\textbf{Input: } A target diagonal matrix $\mathrm{D}(\bm{\theta})$ with Eq. \eqref{intro:eq-theta}. \\
\textbf{Output: } A CNOT-optimized $\{\mathrm{CNOT},R_z\}$-based circuit $\mathrm{QC}$ for fulfilling $\mathrm{D}(\bm{\theta})$ up to a global phase. \\
\hline
\textbf{ 1 } Calculate rotation angles $\varphi_j = - \frac{2}{2^{2n}}\sum_{k=0}^{2^{2n}-1} (-1)^{\sigma(j,k)}\theta_k$, 
for $j=[j_0 j_1 \ldots j_{2n-1}]$ satisfying \\
\hspace{0.5cm} $\sum_{m=0}^{2n-1} j_m \mod 2 = 0$; \\
\textbf{ 2 } $\mathrm{QC}$ $\leftarrow [\,]$, 
$\mathrm{CC}_1 = [0]$, $\mathrm{GC}_1 = [0,1]$, $\mathrm{HGC}_1 = [1]$, $\mathrm{RAI}_1 = [110^{(2n-2)}]$; \\
\textbf{ 3 \ for } $m=1$ \textbf{to } $2n-2$ \textbf{do } \\
\textbf{ 4 }\hspace{0.8cm} $\mathrm{CC}_{m+1} \leftarrow \mathrm{CC}_{m}$; \\
\textbf{ 5 }\hspace{0.8cm} $\mathrm{CC}_{m+1}[2^{m-1}] \leftarrow m$; 
\text{ // Take value $m$ for the $2^{m-1}$-th component (i.e.,\! the last component).} \\
\textbf{ 6 }\hspace{0.8cm} $\mathrm{CC}_{m+1} \leftarrow [\mathrm{CC}_{m+1},\mathrm{CC}_{m+1}]$; \\
\textbf{ 7 }\hspace{0.8cm} $\mathrm{GC}_{m+1} \leftarrow [\mathrm{GC}_{m}\{0\},\overline{\mathrm{GC}_{m}}\{1\}]$; 
\text{ // $\overline{\mathrm{GC}_{m}}$: reverse of $\mathrm{GC}_m$.} \\
\textbf{ 8 }\hspace{0.8cm} $\mathrm{HGC}_{m+1} \leftarrow \mathrm{GC}_{m+1}[2:2:2^m]$; 
\text{ // Take even components in $\mathrm{GC}_{m+1}$ as $\mathrm{HGC}_{m+1}$.} \\
\textbf{ 9 }\hspace{0.8cm} $\mathrm{RAI}_{m+1} \leftarrow \mathrm{HGC}_{m+1}\{1\}\{0^{(2n-2-m)}\}$; \\
\hspace{1.0cm} \text{ // $\mathrm{HGC}_{m+1}\{1\}\{0^{(2n-2-m)}\}$: append $\{1\}$ and $2n-2-m$ times $\{0\}$ after its each component.} \\
\textbf{10 \ end } \\
\textbf{11 \ for } $m=1$ \textbf{to } $2n-1$ \textbf{do } \\
\textbf{12 }\hspace{0.7cm} Append $\mathrm{CNOT}(2n-m,0)$ to $\mathrm{QC}$; \\
\textbf{13 \ end } \\
\textbf{14 \ for } $m=1$ \textbf{to } $2n-1$ \textbf{do } \\
\textbf{15 }\hspace{0.7cm} \textbf{if } $m<2n-1$ \textbf{then } \\
\textbf{16 }\hspace{1.2cm} \textbf{for } $i=1$ \textbf{to } $2^{2n-2-m}$ \textbf{do } \\
\textbf{17 }\hspace{1.9cm} \textbf{for } $j=1$ \textbf{to } $m$ \textbf{do } \\
\textbf{18 }\hspace{2.6cm} index $\leftarrow 2^{2n-2-m+j}-2^{2n-1-m}+i$; \\
\textbf{19 }\hspace{2.6cm} k $\leftarrow \mathrm{RAI}_{2n-1-m+j}[\mathrm{index}]$; \\
\textbf{20 }\hspace{2.6cm} Append $\mathrm{F}(2n-1-m+j,\mathrm{CC}_{2n-1-m+j}[\mathrm{index}],\varphi_{k})$ to $\mathrm{QC}$; \\
\textbf{21 }\hspace{1.9cm} \textbf{end } \\
\textbf{22 }\hspace{1.2cm} \textbf{end } \\
\textbf{23 }\hspace{0.7cm} \textbf{else } \\
\textbf{24 }\hspace{1.2cm} \textbf{for } $j=1$ \textbf{to } $2n-1$ \textbf{do } \\
\textbf{25 }\hspace{1.9cm} index $\leftarrow 2^{j-1}$; \\
\textbf{26 }\hspace{1.9cm} k $\leftarrow \mathrm{RAI}_{j}[\mathrm{index}]$; \\
\textbf{27 }\hspace{1.9cm} Append $\mathrm{F}(j,\mathrm{CC}_{j}[\mathrm{index}],\varphi_{k})$ to $\mathrm{QC}$; \\
\textbf{28 }\hspace{1.2cm} \textbf{end } \\
\textbf{29 }\hspace{0.7cm} \textbf{end } \\
\textbf{30 \ end } \\
\textbf{31 \ return } $\mathrm{QC}$ \\
\hline
\end{tabular}
\end{minipage}
}%
\end{table*}

\subsection{Evaluation of circuit depth}
\label{sec:4-3}

First, we estimate the depth of the circuit using Algorithm 1. 
For a fixed $n\in \mathbb{N}$, the circuit depth can be derived using computer software (e.g.,\! qiskit \cite{Qiskit23} provided by IBM). 
Here, instead of the numerical evaluation, we provide an analytic overhead of the depth. 
To do this, we check whether the $R_z$ gate in each F-block can be executed simultaneously with the previous CNOT gate in the preceding F-block. 

From the structure of the F-blocks, it is easy to find that the $R_z$ gate in an F-block can be embedded into the previous block if and only if the target qubit of this F-block is different from both the control qubit and the target qubit of the preceding F-block or CNOT gate. We check this from Step $0$ to Step $2n-1$: 

\noindent -- Step $0$: In this step, we have only $2n-1$ CNOT gates and there are no $R_z$ gates. 

\noindent -- Step $m$ ($m=1,2,\ldots,2n-2$): In these intermediate steps, we manage to reduce the circuit depth. Under our construction (see especially $\{\mathrm{CC}_m\}_{m=1}^{2n-1}$), we can track the control and target qubits in the F-blocks by the triangle-like arrangement in Table \ref{tab:4}.  
\begin{table}[htb]
\caption{Representation of the target qubits and control qubits of F-blocks in a triangle-like table. The control qubits for different $\mathrm{B}_{k}$ blocks are distinguished using dashed lines. }\label{tab:4}
\tiny\centering
\scalebox{1.2}[1.2]{
\tabcolsep = 0.1cm
\begin{tabular}{c|rrrrrrrr|rrrr|rr|c|c}
\hline
Targ. qubit no. $m$ & \multicolumn{16}{c}{Cont. qubit no. $\mathrm{CC}_m$}\\
\hline
1 &  &  &  &  &  &  &  &  
&  &  &  &  &  &  &  & 0 \\
2 &  &  &  &  &  &  &  &  
&  &  &  &  &  &  & 1 & 1 \\
3 &  &  &  &  &  &  &  &  
&  &  &  &  & 1 & 2 & 1 & 2 \\
4 &  &  &  &  &  &  &  &  
& 1 & 2 & 1 & 3 & 1 & 2 & 1 & 3 \\
5 & 1 & 2 & 1 & 3 & 1 & 2 & 1 & 4 
& 1 & 2 & 1 & 3 & 1 & 2 & 1 & 4 \\
$\cdots$ &\multicolumn{8}{c|}{$\cdots$} & \multicolumn{4}{c|}{$\cdots$} & 
\multicolumn{2}{c|}{$\cdots$} & \multicolumn{1}{c|}{$\cdots$} & 
\multicolumn{1}{c}{$\cdots$} \\
\hline
\end{tabular}
}
\end{table}
Recall that in Step $m$, we have block $\mathrm{B}_{2n-1-m}$.  
The target qubit of the first F-block in $\mathrm{B}_{k}$ coincides with the control qubit of the last F-block in the previous $\mathrm{B}_{k+1}$, $k=1,\ldots,2n-2$. 
Subsequently, the $R_z$ gates in the first F-block in $\mathrm{B}_k$ cannot be embedded into the preceding CNOT gates. 
On the other hand, we have the same control qubits in each column in the representation of $\mathrm{B}_k$, $k=1,2,\ldots,2n-3$ (see Table \ref{tab:4}). 
According to Algorithm 1, the target qubit nos. in each column are strictly larger than the control qubit nos. of all the F-blocks in this column, which implies that we can save a depth of $k-1$ (where $k$ is the number of F-blocks) in each column. 
Moreover, for $m=2,3,\ldots,2n-2$, the $R_z$ gate of the first F-block in each column can be embedded into the previous F-block in the previous column because their target qubits are different. 
Thus, we can save $k_m-1$ depth in each $\mathrm{B}_{2n-1-m}$ block, where $k_m = m 2^{2n-2-m}$ is the number of F-blocks in $\mathrm{B}_{2n-1-m}$, $m=2,3,\ldots,2n-2$. 
Direct observation shows that we also save $1$ depth in $\mathrm{B}_{2n-2}$ (i.e.,\! the $R_z$ gate in the first F-block).  

\noindent -- Step $2n-1$: In this step, we have $2n-1$ F-blocks and find the target qubit no. of each F-block is one larger than the target qubit no. and two larger than the control qubit no. of the previous F-block, which indicates that we can save $2n-2$ depth in this step. 

From the above discussion, we find that we save (at least) 
$2^{2n-1}-1-(2^{2n-3}-1)-(2n-2) = 3\times 2^{2n-3}-2n+2$ depth for $n\ge 2$. Therefore, let $d_n$ be the depth of the circuit generated by Algorithm 1, then we have 
\begin{align*}
d_n &\le 2n-1 +2(2^{2n-1}-1) -(3\times 2^{2n-3}-2n+2) \\
&= 2^{2n-1} + 2^{2n-3} + 4n-5 \quad \text{ for } n\ge 2,
\end{align*}
and $d_1 = 3$. This upper bound is nearly halved compared to the circuit depth $2^{2n}$ derived for general diagonal unitary matrices.   


\subsection{Further reduction in circuit depth}
\label{sec:4-4}

Intrigued by Fig.~\ref{pre:fig3} in Sect.~\ref{sec:3} (or more precisely, Fig.~\ref{app:figsim3} in Appendix \ref{app:subsec-2}), we discuss the further reduction of circuit depth by applying the rewriting rules (b), (c) in Fig.~\ref{pre:fig2}. For simplicity, we demonstrate this reduction in the example of $n=2$. 

According to our proposed algorithm, we first write the generated circuit by the CNOT gates and F-blocks in the following order: 
{\small
$$
\begin{aligned}
&\mathrm{CNOT}(3,0), \mathrm{CNOT}(2,0), \mathrm{CNOT}(1,0), 
\mathrm{F}(3,1,\varphi_{1001}), \mathrm{F}(3,2,\varphi_{0101}),\\
&\mathrm{F}(2,1,\varphi_{1010}), \mathrm{F}(3,1,\varphi_{1111}), \mathrm{F}(1,0,\varphi_{1100}), \mathrm{F}(2,1,\varphi_{0110}), 
\mathrm{F}(3,2,\varphi_{0011}). 
\end{aligned}
$$
}

\noindent Next, we write each F-block into an $R_z$ gate and a CNOT gate, and by combining the above discussion, we obtain
{\small
$$
\begin{aligned}
&\mathrm{CNOT}(3,0), \mathrm{CNOT}(2,0), \mathrm{CNOT}(1,0)R_z(3,\varphi_{1001}),\mathrm{CNOT}(3,1), \\
&R_z(3,\varphi_{0101}), \mathrm{CNOT}(3,2), R_z(2,\varphi_{1010}),\mathrm{CNOT}(2,1)R_z(3,\varphi_{1111}), \\
&\mathrm{CNOT}(3,1), R_z(1,\varphi_{1100}), \mathrm{CNOT}(1,0)R_z(2,\varphi_{0110}), \\
&\mathrm{CNOT}(2,1)R_z(3,\varphi_{0011}),\mathrm{CNOT}(3,2), 
\end{aligned}
$$
}

\noindent with depth $13$. The depth can be obtained by counting the commas in the above expression because the gates between every two commas can be executed simultaneously. 
Moreover, by applying the rewriting rule (c) in Fig.~\ref{pre:fig2} we move the $R_z$ gate in the first F-block of $\mathrm{B}_1,\mathrm{B}_0$ (i.e.,\! from $\mathrm{B}_{2n-3}$ to $\mathrm{B}_0$ in general) to the place of the $R_z$ gate in the second F-block in $\mathrm{B}_2$ (i.e.,\! $\mathrm{B}_{2n-2}$ in general) to obtain 
{\small
$$
\begin{aligned}
&\mathrm{CNOT}(3,0), \mathrm{CNOT}(2,0), \mathrm{CNOT}(1,0)R_z(3,\varphi_{1001}),\mathrm{CNOT}(3,1),\\
&R_z(3,\varphi_{0101})R_z(2,\varphi_{1010})R_z(1,\varphi_{1100}),
\mathrm{CNOT}(3,2), \\
&\mathrm{CNOT}(2,1)R_z(3,\varphi_{1111}), \mathrm{CNOT}(3,1), \mathrm{CNOT}(1,0)R_z(2,\varphi_{0110}), \\
&\mathrm{CNOT}(2,1)R_z(3,\varphi_{0011}),\mathrm{CNOT}(3,2), 
\end{aligned}
$$
}

\noindent with depth $11$. 
Finally, using the rewriting rule (b), we move $\mathrm{CNOT}(2,0)$ (i.e.,\! $\mathrm{CNOT}(k,0)$, $k=2,3,\ldots,2n-2$ in general) into the first F-block in $\mathrm{B}_2$ and then we arrive at
{\small
$$
\begin{aligned}
&\mathrm{CNOT}(3,0), \mathrm{CNOT}(1,0)R_z(3,\varphi_{1001}),\mathrm{CNOT}(2,0)\mathrm{CNOT}(3,1),\\
&R_z(3,\varphi_{0101})R_z(2,\varphi_{1010})R_z(1,\varphi_{1100}), \mathrm{CNOT}(3,2), \\
&\mathrm{CNOT}(2,1)R_z(3,\varphi_{1111}), \mathrm{CNOT}(3,1), \mathrm{CNOT}(1,0)R_z(2,\varphi_{0110}), \\
&\mathrm{CNOT}(2,1)R_z(3,\varphi_{0011}),\mathrm{CNOT}(3,2), 
\end{aligned}
$$
}

\noindent with depth $10$, which is exactly the circuit derived in Fig.~\ref{pre:fig3}. 

Therefore, for general $n\ge 2$, we can further reduce the depth by $2n-2+2n-3=4n-5$ with the help of the rewriting rules (b), (c) in Fig.~\ref{pre:fig2}, and hence we obtain the upper bound of the circuit depth:  
$$
d_n \le 2^{2n-1} + 2^{2n-3}. 
$$

\section{Applications to Hamiltonian Simulation}
\label{sec:appl}

In this section, we apply Algorithm 1 to Hamiltonian simulation in one dimension 
and show that our algorithm is effective and has promising applications in some quantum algorithms/subroutines. 
The numerical simulation results are based on the quantum gate simulation 
using qiskit \cite{Qiskit23}. 

For the first quantized Hamiltonian simulation, there is a well-known grid-based method that uses the so-called centered/shifted quantum Fourier transform $U_{\text{CQFT}}$ to diagonalize the kinetic energy operator $\hat{T}$. By noting that the potential energy operator $\hat{V}$ itself is already diagonal in the real-space (position) representation, we combine this with the Trotter-Suzuki formula to obtain an approximation scheme \cite{Kassal.08, Childs.22, Chan.23, Kosugi.23}. 
For example, the first-order Trotter-Suzuki formula gives the following approximation for $K$ steps:
\begin{align*}
e^{-\mathrm{i}\mathcal{H}K\Delta t} &= (e^{-\mathrm{i}\mathcal{H}\Delta t})^K \\
&\approx (e^{-\mathrm{i}\hat{T}\Delta t} e^{-\mathrm{i}\hat{V}\Delta t})^K \\
&= \left(U_{\text{CQFT}} U_{\text{kin}}(\Delta t)U_{\text{CQFT}}^{\dag} U_{\text{pot}}(\Delta t)\right)^K,
\end{align*}
where $U_{\text{kin}}, U_{\text{pot}}$ are two diagonal unitary matrices with the parameter $\Delta t$. 

\subsection{One-particle simulation with Eckart barrier potential}

First, we consider a Hamiltonian simulation with one electron, in which we assume the presence of an Eckart barrier. More precisely, we investigate the following Hamiltonian:
\begin{align*}
\mathcal{H} = \hat{T} + \hat{V} = \frac{p_e^2}{2m_e} + v_{\text{eck}}(|r_e-r_0|). 
\end{align*}
Here, $r_e$ and $p_e$ denote the position and momentum of an electron, respectively. 
$\hat{T}$ is the kinetic energy operator of an electron having the mass $m_e=1$, 
while $\hat{V}$ consists of an Eckart barrier potential in the form of 
$v_{\text{eck}}(x) = A\, \mathrm{sech}(ax)$ with two parameters $A$ and $a$ 
indicating the strength and sharpness of the barrier, and $r_0$ is the center point of the potential. 
Since the Eckart potential is an even function, we can use Algorithm 1 as a subroutine 
for the implementation of the potential part. 
Indeed, we assume $n=2n^\prime$ to fit the setting in the above sections (for odd $n$, we can also apply Algorithm 1 by substituting $2n$ with $n$). Let $L$ be the length of the simulation cell and $x_k = k\Delta x - L/2$, $k=0,1,\ldots,2^n-1$ be the grid points, where $\Delta x=L/2^n$. Then we find 
\begin{align}
\label{intro:eq-theta3}
& \theta_{j,j^\prime} := -v_{\text{eck}}\left(\left|2^{n^\prime} j+j^\prime-2^{n-1}+1/2\right|\Delta x\right)\Delta t, 
\end{align}
for $j,j^\prime = 0,1,\ldots, 2^{n^\prime}-1$, satisfies the assumption \eqref{intro:eq-theta}, and thus the corresponding diagonal unitary matrix $\mathrm{D}(\bm{\theta})$ implemented by Algorithm 1 is equivalent to the real-time evolution operator $e^{-\mathrm{i}\hat V\Delta t}$ with $\hat V = \sum_{k=0}^{2^n-1}v_{\text{eck}}(|x_k-r_0|)\ket{k}\bra{k}$ where $r_0 = (-1/2)\Delta x = -L/2^{n+1}$. 

For the kinetic part, one can employ the subroutine either by a general diagonal unitary matrix \cite{ZHL22} or by a polynomial phase gate \cite{Ollitrault.20, Kosugi.23} (as the kinetic energy is a quadratic function of the momentum). 
\begin{figure*}[htb]
\centering
\scalebox{0.36}[0.36]{
\includegraphics[keepaspectratio]{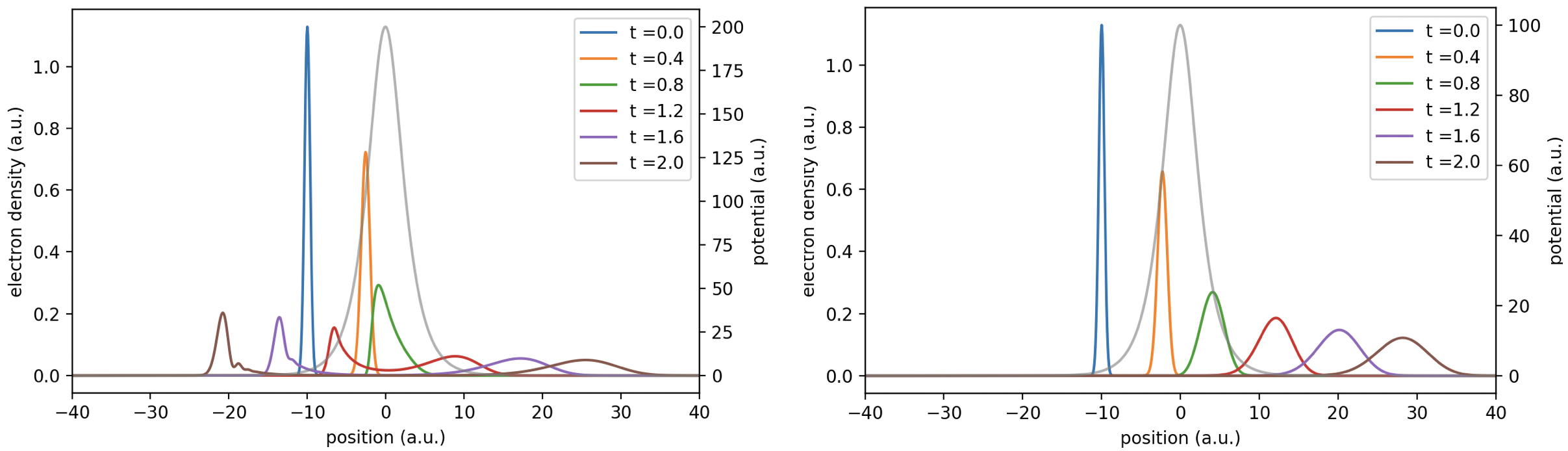}
}%
\caption{(Color online) Hamiltonian simulation with different strength of the barrier at time points $t=0, 0.4, 0.8, 1.2, 1.6, 2.0$ with grid size $n=10$ and time step $\Delta t = 0.1$. 
In each panel, the left vertical axis represents the electron density, and the right vertical axis represents the Eckart barrier potential in gray. 
The left panel shows the case with a larger strength $A=200$ where one observes that a partial wave penetrates the barrier while a partial wave returns. The right panel shows the case with a weaker strength $A=100$ in which only penetration is observed. }
\label{appl:fig1}
\end{figure*}
We consider a Gaussian wave packet, and to simulate the behavior when the wave packet interacts with the barrier, we set the initial state as
\begin{align*}
\ket{\psi(x, 0)} \propto e^{-\frac{(x-x_0)^2}{2\sigma^2}+\mathrm{i}k_0(x-x_0)}, 
\end{align*}
where $x_0$ and $\sigma$ denote the mean and the standard deviation of the Gaussian distribution, respectively, and $k_0$ is a momentum parameter. 
For fixed $k_0=20$, $\sigma=0.5$, $a=0.5$, $L=80$ and $x_0=-10$, 
we observe in Fig.~\ref{appl:fig1} that the wave packet penetrates the barrier with a weak strength $A=100$ and partially penetrates the barrier with a larger strength $A=200$. 
Here, we choose $n=10$ and $\Delta t = 0.1$ for the simulation, and we find that the infidelity between the simulated wave function using a quantum emulator and the ``exact" one (calculated by diagonalizing the discretized Hamiltonian matrix) is less than $0.05$. 

Although we show the simulation result only for $n=10$ as an example, we can readily estimate the CNOT count and circuit depth for the Hamiltonian simulation in each step with a couple of $n$'s. The depth is derived by counting single-qubit gates and CNOT gates that cannot be executed simultaneously. 
Moreover, we are interested in a comparison with the circuit in Ref. \cite{ZHL22}. The results are listed in Tables~\ref{tab:app1} and \ref{tab:app1b}, where we use the polynomial phase gate \cite{Kosugi.23} for the kinetic part. 
\begin{table*}[htb]
\centering
\caption{CNOT count for the one-particle Hamiltonian simulation for a single Trotter step with first-order Trotter-Suzuki formula. }
\label{tab:app1}
\scalebox{1}[1]{
\begin{tabular}{l|ccccccc|l}
\hline
CNOT count & $n=7$ & $n=8$ & $n=9$ & $n=10$ & $n=11$ & $n=12$ & $n=13$ & $n\ge 2$\\
\hline
Potential part (Algorithm 1) & 69 & 134 & 263 & 520 & 1033 & 2058 & 4107 & $2^{n-1}+n-2$\\
Potential part (Ref. \cite{ZHL22}) & 126 & 254 & 510 & 1022 & 2046 & 4094 & 8190 & $2^{n}-2$\\
Total (Algorithm 1+Ref. \cite{Kosugi.23})& 195 & 302 & 479 & 790 & 1363 & 2454 & 4575 & $2^{n-1}+3n^2-2n-2$\\
Total (Ref. \cite{ZHL22}+Ref. \cite{Kosugi.23}) & 252 & 422 & 726 & 1292 & 2376 & 4490 & 8658 & $2^{n}+3n^2-3n-2$\\
\hline
\end{tabular}
}
\end{table*}
\begin{table*}[htb]
\centering
\caption{Circuit depth for the one-particle Hamiltonian simulation for a single Trotter step with first-order Trotter-Suzuki formula. }
\label{tab:app1b}
\scalebox{1}[1]{
\begin{tabular}{l|ccccccc|l}
\hline
Depth & $n=7$ & $n=8$ & $n=9$ & $n=10$ & $n=11$ & $n=12$ & $n=13$ & $n\ge 2$\\
\hline
Potential part (Algorithm 1) & 80 & 160 & 320 & 640 & 1280 & 2560 & 5120 & $2^{n-1}+2^{n-3}$\\
Potential part (Ref. \cite{ZHL22}) & 127 & 255 & 511 & 1023 & 2047 & 4095 & 8191 & $2^{n}-1$\\
Total (Algorithm 1+Ref. \cite{Kosugi.23}) & 202 & 302 & 482 & 822 & 1482 & 2782 & 5362 & $2^{n-1}+2^{n-3}+20n-18$\\
Total (Ref. \cite{ZHL22}+Ref. \cite{Kosugi.23}) & 249 & 397 & 673 & 1205 & 2249 & 4317 & 8433 & $2^{n}+20n-19$\\
\hline
\end{tabular}
}
\end{table*}
It is clear that Algorithm 1 provides a shallower circuit than directly using the result 
\cite{ZHL22}, and the reduction of CNOT count and depth for the even potential part 
is asymptotically $50\%$. 
Furthermore, since the kinetic part is implemented by a polynomial phase gate, which has a CNOT count of order $\mathcal{O}(n^2)$ and a depth of $\mathcal{O}(n)$, the dominant resource required for large $n$ depends on the potential part. Thus, we also find about $39\%$ reduction in the CNOT count and $32\%$ reduction in the depth for a practical number $n=10$.   

\subsection{Two-particle simulation in an electric field}

Next, we discuss a Hamiltonian simulation with two electrons in a known static electric field. 
In detail, we consider a LiH molecule simulation where the two electrons 
are bound to the corresponding Li or H nucleus separately at the beginning. 
From $t=0$, we impose an electric field and demonstrate the evolution of the two-electron system 
by using a quantum emulator. 

Under the above setting, we introduce the following Hamiltonian:
\begin{align*}
\mathcal{H} &= \hat{T} + \hat{V}_{\text{en}} + \hat{V}_{\text{ee}} + \hat{V}_{\text{nn}} + \hat{V}_{\text{ext}} \\
&= \sum_{\ell=1}^2 \frac{p_\ell^2}{2m_\ell} - \sum_{\ell=1}^2 \left(Z_{\text{Li}} v_{\text{eLi}}(|r_\ell-R_{\text{Li}}|) + Z_{\text{H}} v_{\text{eH}}(|r_\ell-R_{\text{H}}|) \right) \\
&\quad + v_{\text{ee}}(|r_1-r_2|) + Z_{\text{Li}} Z_{\text{H}} v_{\text{LiH}} + \sum_{\ell=1}^2 v_{\text{ext}}(r_\ell).  
\end{align*}
\noindent Here, we regard the nucleus as fixed classical particles and consider the system of two electrons. 
$r_\ell$ and $p_\ell$ denote the position and momentum of the $\ell$-th electron, respectively, and the mass of the electrons and the charge of the nuclei are set to be unit: $m_\ell=1$, $Z_{\text{Li}} = Z_{\text{H}} = 1$. 
Moreover, the particle-particle interaction is modelled by modified Coulomb potential 
$v_{X}(r) = 1/{\sqrt{\lambda_{X}^2+r^2}}$ for $X = \text{eH}, \text{eLi}, \text{ee}, \text{LiH}$ 
with parameters $\lambda_{\text{eH}}^2 = 0.7, \lambda_{\text{eLi}}^2 = 2.25, \lambda_{\text{ee}}^2 = 0.6$ and  $\lambda_{\text{LiH}}^2 = \lambda_{\text{eH}}^2 + \lambda_{\text{eLi}}^2 - \lambda_{\text{ee}}^2 = 2.35$ which we take from Ref. \cite{Kosugi.23} as model parameters. 
Furthermore, the external potential originating from the static electric field is described by 
$v_{\text{ext}}(x) = - \omega_0 x$ where the parameter $\omega_0$ indicates the strength of the electric field. 
Let $L$ be the length of the simulation cell and $x_k = k\Delta x - L/2$, $k=0,1,\ldots,2^n-1$ be the grid points, where $\Delta x=L/2^n$. Then we find 
\begin{equation}
\label{intro:eq-theta2}
\theta_{j,j^\prime} := -v_{\text{ee}}(|j-j^\prime|\Delta x)\Delta t, \quad j,j^\prime = 0,1,\ldots, 2^n-1,
\end{equation}

\noindent satisfies the assumption \eqref{intro:eq-theta} (see the verification before Corollary 1 
in Appendix \ref{app:subsec-1}). Then the corresponding diagonal unitary matrix $\mathrm{D}(\bm{\theta})$ is equivalent to the real-time evolution operator $e^{-\mathrm{i}\hat V_{ee}\Delta t}$, which is the part of electron-electron interaction. The kinetic part can be again implemented by the phase gate for polynomial separately for each electron, and the rest electron-nucleus and nucleus-nucleus part can be implemented using general diagonal unitary matrices \cite{ZHL22}, which is efficient for small grid parameter $n$. 
\begin{figure*}[htb]
\centering
\scalebox{0.36}[0.36]{
\includegraphics[keepaspectratio]{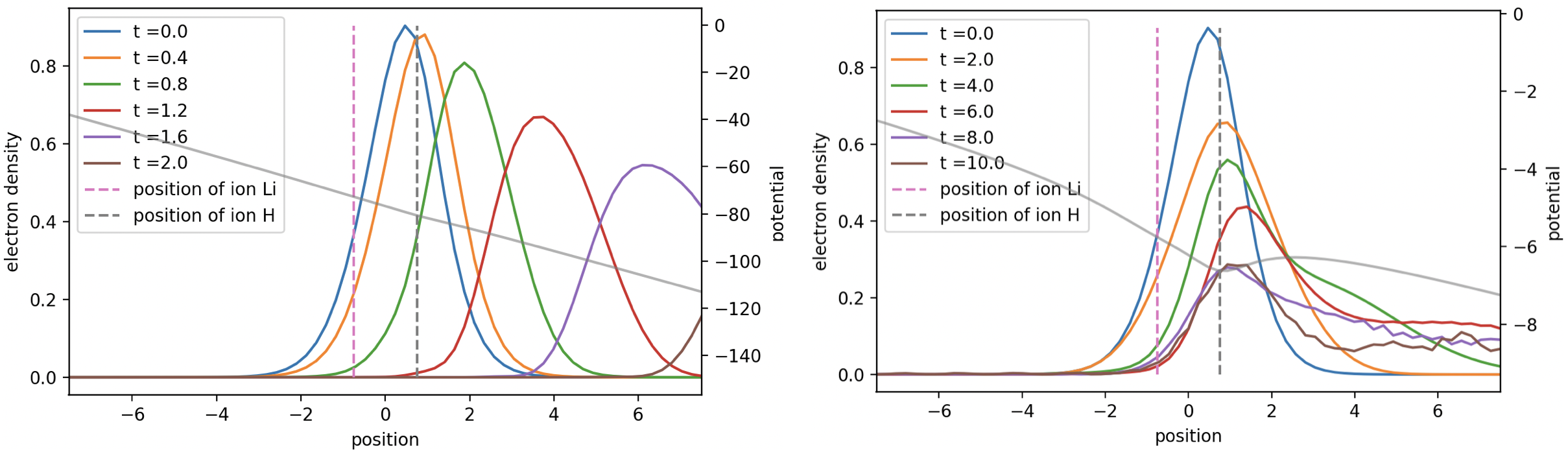}
}%
\caption{(Color online) Hamiltonian simulation with grid size $n=7$, time step $\Delta t = 0.1$, and different electric fields. 
In each panel, the left vertical axis represents the electron density of the two-electron system, and the right vertical axis represents the potential due to nuclei in gray. 
The left panel shows the fast decay of the electron density in the virtual domain of the molecule with a strong electric field of parameter $\omega_0=5$, and the probability of electrons remaining in the molecule is very small after $t>2.0$. 
The right panel demonstrates the case with a weak electric field of parameter $\omega_0=0.3$, in which we observe that the electrons remain in the molecule with high probability even at $t=10.0$. }
\label{appl:fig2}
\end{figure*}
As for this simulation, we set the initial condition as the ground state for the LiH molecule model without the electric field, which is derived with classical computation by calculating the eigenstate of the discretized Hamiltonian matrix for the smallest eigenvalue. 
The ground state preparation can be done using the state loading subroutines with different techniques \cite{Ollitrault.20, Rattew.21, Rattew.22}. Here, we do not focus on this issue, and precisely load the calculated state vector into the quantum emulator. 
Moreover, we set $R_{\text{H}}=0.75$, $R_{\text{Li}}=-0.75$ and $L=30$, and the result of the time evolution of the wave function (the electron density) is shown in Fig.~\ref{appl:fig2}. 
As we observe, the electron density in the interval $[-7.5, 7.5]$ (i.e., virtual domain of the LiH molecule) almost vanishes after $t>2.0$ for a strong electric field with $\omega_0 = 5$, while it even remains at $t=10.0$ for the case of a weak electric field with $\omega_0 = 0.3$. 
We choose $n=7$ and $\Delta t = 0.1$ for the simulation, and the infidelity between the simulated wave function and the ``exact" one (calculated by diagonalizing the discretized Hamiltonian matrix) is less than $0.001$ for the case of a strong electric field, and is less than $0.05$ for the case of a weak electric field. 
Besides, we mention that the error of the simulation mainly comes from the Trotter splitting, and for all of the numerical plots, we employed the second-order Trotter-Suzuki formula so that we can derive a relatively high fidelity with a moderately small time step $\Delta t$.  

We estimate the CNOT count and circuit depth for the Hamiltonian simulation in each step with a couple of $n$'s and compare it with a known circuit. The results are listed in Tables~\ref{tab:app2} and \ref{tab:app2b} where we use the polynomial phase gate for the kinetic part and Ref. \cite{ZHL22} for the potential part except for the electron-electron interaction. 
\begin{table*}[htb]
\centering
\caption{CNOT count for the two-particle Hamiltonian simulation for a single Trotter step with the first-order Trotter-Suzuki formula. }
\label{tab:app2}
\scalebox{1}[1]{
\begin{tabular}{l|cccccc|l}
\hline
CNOT count & $n=4$ & $n=5$ & $n=6$ & $n=7$ & $n=8$ & $n=9$ & $n\ge 2$\\
\hline
E.-e. interaction part (Algorithm 1) & 134 & 520 & 2058 & 8204 & 32782 & 131088 & $2^{2n-1}+2n-2$\\
E.-e. interaction part (Ref. \cite{ZHL22}) & 254 & 1022 & 4094 & 16382 & 65534 & 262142 & $2^{2n}-2$\\
Total (Algorithm 1+Ref. \cite{ZHL22}+Ref. \cite{Kosugi.23}) & 234 & 700 & 2362 & 8708 & 33626 & 132540 & $2^{2n-1}+2^{n+1}+6n^2-4n-6$\\
Total (Ref. \cite{ZHL22}+Ref. \cite{Kosugi.23}) & 354 & 1202 & 4398 & 16886 & 66378 & 263594 & $2^{2n}+2^{n+1}+6n^2-6n-6$\\
\hline
\end{tabular}
}
\end{table*}
\begin{table*}[htb]
\centering
\caption{Circuit depth for the two-particle Hamiltonian simulation for a single Trotter step with the first-order Trotter-Suzuki formula. }
\label{tab:app2b}
\scalebox{1}[1]{
\begin{tabular}{l|cccccc|l}
\hline
Depth & $n=4$ & $n=5$ & $n=6$ & $n=7$ & $n=8$ & $n=9$ & $n\ge 2$\\
\hline
E.-e. interaction part (Algorithm 1) & 160 & 640 & 2560 & 10240 & 40960 & 163840 & $2^{2n-1}+2^{2n-3}$\\
E.-e. interaction part (Ref. \cite{ZHL22}) & 255 & 1023 & 4095 & 16383 & 65535 & 262143 & $2^{2n}-1$\\
Total (Algorithm 1+Ref. \cite{ZHL22}+Ref. \cite{Kosugi.23}) & 237 & 753 & 2725 & 10489 & 41357 & 164513 & $2^{2n-1}+2^{2n-3}+2^n+20n-19$\\
Total (Ref. \cite{ZHL22}+Ref. \cite{Kosugi.23}) & 332 & 1136 & 4260 & 16632 & 65932 & 262816 & $2^{2n}+2^n+20n-20$\\
\hline
\end{tabular}
}%
\end{table*}
For the case of $n=7$, Algorithm 1 gives about $48\%$ reduction in the CNOT count and $37\%$ reduction in the depth for the implementation of the real-time evolution operator compared to that using only Ref. \cite{ZHL22}. 

\subsection{Discussion} 

We close this section with a possibly more efficient implementation of the interaction potential using one ancillary qubit. 
The idea is to construct a distance register $\ket{|j-j^\prime|}_n$, and hence the implementation is reduced to a one-variable function depending only on the distance. Such an idea has already been discussed (even for high space dimension) in some previous works on Hamiltonian simulation, but we do not find any result on the detailed gate count of such a circuit. 
Here, we provide a detailed gate implementation and discuss the CNOT count and the circuit depth for different numbers of qubits. 

\begin{figure}
\centering
\resizebox{13.0cm}{!}{
\includegraphics[keepaspectratio]{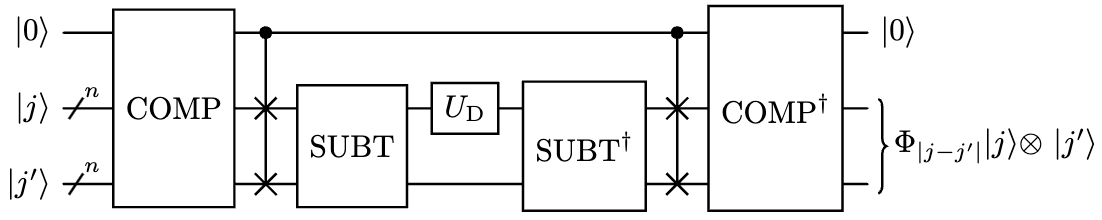}
}%
\caption{Proposed quantum circuit by combining Ref. \cite{ZHL22} with quantum comparator COMP and modular subtractor SUBT. Here, $U_{\mathrm{D}}$ is the diagonal unitary operation $\ket{k} \rightarrow e^{-\mathrm{i}V(k\Delta x)\Delta t}\ket{k}$, $k=0,1,\ldots,2^n-1$ and the final state $\Phi_{|j-j^\prime|} \ket{j}\otimes \ket{j^\prime}$ is $e^{-\mathrm{i}V(|j-j^\prime|\Delta x)\Delta t} \ket{j}\otimes \ket{j^\prime}$, both up to a global phase and we suggest applying the quantum comparator/modular subtractor \cite{Li.20} (details are given in Appendix \ref{app:subsec-4}). }
\label{dis:fig1}
\end{figure}

In one dimension, the distance register can be derived by using a quantum comparator, a Fredkin gate, and a modular subtractor, and then the quantum circuit can be constructed as shown in Fig.~\ref{dis:fig1}. For the one-variable function, we employ the circuit in Ref. \cite{ZHL22}, while this time we apply it to an $n$-qubit distance register instead of the original $2n$-qubit position register. 
This proposed circuit involves the uncomputation part, which enables one to discard or reuse the ancillary qubit safely by disentangling it from the other qubits.
By using the comparator and modular subtractor in Ref. \cite{Li.20} (see also Appendix \ref{app:subsec-4}), which are the current best ones (in gate count) to the authors' knowledge, we have the comparison between Algorithm 1 and the circuit by Fig.~\ref{dis:fig1}, as shown in Table \ref{tab:6}. 
\begin{table*}[htb]
\centering
\caption{Comparison between Algorithm 1 and Fig.~\ref{dis:fig1}. The values in brackets in the last two rows are the depths obtained by numerical tests using qiskit \cite{Qiskit23}. }
\label{tab:6}
\scalebox{1}[1]{
\begin{tabular}{l|ccccccc|l}
\hline
& $n=1$ & $n=2$ & $n=3$ & $n=4$ & $n=5$ & $n=6$ & $n=7$ & $n\ge 3$\\
\hline
CNOT count (Algorithm 1) & 2 & 10 & 36 & 134 & 520 & 2058 & 8204 & $2^{2n-1}+2n-2$\\
CNOT count (Fig.~\ref{dis:fig1}) & - & 92 & 176 & 266 & 364 & 478 & 624 & $2^{n}+82n-78$\\
Depth (Algorithm 1) & 3(3) & 10(10) & 40(39) & 160(155) & 640(561) & 2560(2155) & 10240(8501) & $2^{2n-1}+2^{2n-3}$\\
Depth (Fig.~\ref{dis:fig1}) & - & 134(108) & 244(175) & 356(250) & 476(332) & 612(430) & 780(560) & $2^{n}+104n-76$\\
\hline
\end{tabular}
}%
\end{table*}
For small $n$, Algorithm 1 gives a smaller and shallower implementation, while for $n\ge 5$, Fig.~\ref{dis:fig1} achieves lower cost in both the CNOT count and the depth due to its slower increasing with respect to $n$. 

\section{Conclusion}
\label{sec:6}

In this paper, we discussed the construction of a quantum circuit for implementing a specific type of symmetric diagonal unitary matrix $\mathrm{D}(\bm{\theta})$ with reflection symmetry that satisfies Eq. \eqref{intro:eq-theta}, based on $\{\mathrm{CNOT}, R_z\}$ gates. 
Such diagonal unitary matrices can be used to implement the real-time evolution of the interaction potential part or some even potential functions in the first quantized Hamiltonian simulation. 
Owing to such a symmetric structure, we achieved by Algorithm 1 nearly half the reduction in both the number of entanglement gates and circuit depth compared to the existing optimal constructive algorithm for general diagonal unitary matrices \cite{ZHL22} (see Tables \ref{tab:app1}--\ref{tab:app2b}). 

Although we conjecture that the number of entanglement gates is at best $2^{2n-1}+2n-2$ in the $\{\mathrm{CNOT}, R_z\}$-based circuit for the worst scenario of diagonal unitary matrices that satisfy Eq. \eqref{intro:eq-theta}, we find by numerical tests (with qiskit) that the circuit depth (based on the CNOT, $R_z$ implementation) can be further reduced by putting some CNOT gates with different control and target qubits together (see the numerical depth in the brackets in Table \ref{tab:6}). 
Therefore, $2^{2n-1}+2^{2n-3}$ is only the upper bound of the depth obtained using our proposed algorithm (i.e.,\! Algorithm 1 with the reduction argument in Sect.~\ref{sec:4-4}). 
Moreover, the theoretical lower bound of the depth can be derived by counting only the F-blocks targeted at $\ket{q_{2n-1}}$, that is, $d_n\ge 2^{2n-1}$. The optimal circuit depth between the theoretical upper and lower bounds
$$
2^{2n-1} \le d_n \le 2^{2n-1} + 2^{2n-3},
$$
is still unknown, and will be the subject of future work. 

Another problem is the efficient quantum circuit for the diagonal matrix that satisfies Eq. \eqref{intro:eq-theta2} in the case of higher dimensions. 
One can directly verify that by suitably arranging the qubits to yield diagonal unitary matrices, Eq. \eqref{intro:eq-theta} still holds for the case of interaction in the first quantized Hamiltonian operators for higher space dimensions. 
For example, for a $3$-dimensional case, $\bm{\theta}$ should be defined by
{\small
\begin{equation}
\label{con:eq1}
\theta_{j_x,j_y,j_z,j_x^\prime,j_y^\prime,j_z^\prime} := - v_{\text{ee}}\left(\sqrt{|j_x-j_x^\prime|^2+|j_y-j_y^\prime|^2+|j_z-j_z^\prime|^2}\Delta x\right)\Delta t,
\end{equation}
}

\noindent instead of Eq. \eqref{intro:eq-theta2}.  
The proposed algorithm then obtains a circuit with $2^{6n-1}+6n-2$ CNOT gates and (at most) depth $2^{6n-1}+2^{6n-3}$. 
In this case, the total number of qubits is $6n$, where $n$ is the grid parameter for each space dimension. However, we find through numerical tests that the rotation gates are further reduced because of the higher-order symmetric property of $\bm{\theta}$ in Eq. \eqref{con:eq1}. 
Consequently, it is expected that the CNOT gate count and depth would be further reduced, which indicates the possibility of a more efficient algorithm to deal with the evolution of the interaction potential in the first quantized Hamiltonians.

\section*{Acknowledgment}
This work was supported by Japan Society for the Promotion of Science (JSPS) KAKENHI under Grant-in-Aid for Scientific Research (Grant numbers 21H04553, 20H00340, and 22H01517) and by Japan Science and Technology Agency (JST) (Grant number JPMJPF2221). 
This study was partially carried out using the TSUBAME3.0 supercomputer at Tokyo Institute of Technology and the facilities of the Supercomputer Center, the Institute for Solid State Physics, The University of Tokyo. 
\appendix
\section{Vanishing of rotation gates}
\label{app:subsec-1}

We give the detailed statement and proof to support our observation that half of $R_z$ gates vanish under the symmetry assumption \eqref{intro:eq-theta}. 
Here, $[j_0 j_1\cdots j_{n-1}]$, $[k_0 k_1\cdots k_{n-1}]$ denote the binary expressions for $j,k$. 
By noting that $\bm\alpha = \mathbf{H}\bm\theta$ can be rewritten in components as follows:
{\small
\begin{equation}
\label{app:eq-alpha}
\alpha_{j,j^\prime} = \frac{1}{\sqrt{2^{2n}}} \sum_{k,k^\prime=0}^{2^n-1} (-1)^{\sigma(j,k)+\sigma(j^\prime,k^\prime)} \theta_{k,k^\prime}, \ j,j^\prime = 0,1,\ldots, 2^n-1,
\end{equation}
}

\noindent where $\sigma(j,k) := j_0k_0 + \cdots + j_{n-1}k_{n-1}$ for $j = [j_0 j_1\cdots j_{n-1}],\, k = [k_0 k_1\cdots k_{n-1}]$, we have the following theorem. 

\noindent {\textbf{Theorem 1. }}
Let $n\in \mathbb{N} := \{1,2,\ldots\}$. 
Assume that $\theta_{k,k^\prime}$, $k,k^\prime = 0,1,\ldots,2^n-1$ satisfies
{\small
$$
\theta_{k,k^\prime} = \theta_{2^n-1-k,2^n-1-k^\prime}, \quad k,k^\prime = 0,1,\ldots,2^n-1. 
$$
}

\noindent Then $\alpha_{j,j^\prime}$ fulfills 
{\small
$$
\alpha_{j,j^\prime} = 0 \quad \text{for any $j,j^\prime$ such that }\quad \sum_{m=0}^{n-1} (j_m + j^\prime_m) \mod 2 = 1. 
$$} 

\noindent {\textbf{Proof. }}
According to the definition of $\alpha_{j,j^\prime}$ and the assumption on $\theta_{k,k^\prime}$, we have
{\small
\begin{align*}
\alpha_{j,j^\prime} 
&= \frac{1}{\sqrt{2^{2n}}} \sum_{k,k^\prime=0}^{2^n-1} (-1)^{\sigma(j,k)+\sigma(j^\prime,k^\prime)} \theta_{k,k^\prime}\\
&= \frac{1}{2\sqrt{2^{2n}}} \sum_{k,k^\prime=0}^{2^n-1} (-1)^{\sigma(j,k)+\sigma(j^\prime,k^\prime)} \theta_{k,k^\prime}\\
&\quad + \frac{1}{2\sqrt{2^{2n}}} \sum_{k,k^\prime=0}^{2^n-1} (-1)^{\sigma(j,2^n-1-k)+\sigma(j^\prime,2^n-1-k^\prime)} \theta_{2^n-1-k,2^n-1-k^\prime}\\
&= \frac{1}{2\sqrt{2^{2n}}} \sum_{k,k^\prime=0}^{2^n-1} \left((-1)^{\sigma(j,k)+\sigma(j^\prime,k^\prime)}+(-1)^{\sigma(j,2^n-1-k)+\sigma(j^\prime,2^n-1-k^\prime)}\right) \theta_{k,k^\prime}\\
&= \frac{1}{2\sqrt{2^{2n}}} \sum_{k,k^\prime=0}^{2^n-1} \left(1+(-1)^{\sigma(j,k)+\sigma(j^\prime,k^\prime)+\sigma(j,2^n-1-k)+\sigma(j^\prime,2^n-1-k^\prime)}\right) \\
&\quad \times (-1)^{\sigma(j,k)+\sigma(j^\prime,k^\prime)}\theta_{k,k^\prime},
\end{align*}
}

\noindent for any $j,j^\prime=0,1,\ldots,2^n-1$. 
Let $p,q = 0,1,\ldots,2^n-1$ admit the binary expression
{\small
$$
p = [p_0 p_1 \ldots p_{n-1}],\quad q = [q_0 q_1 \ldots q_{n-1}].  
$$
}

\noindent We observe that
{\small
\begin{equation}
\label{app:eq-1}
p_m+q_m = 1\quad \forall m=0,1,\ldots,n-1, \quad \text{if $p+q = 2^n-1$}.  
\end{equation}
}

\noindent By using Eq. \eqref{app:eq-1} and the binary expressions of $j,j^\prime,k,k^\prime$, we calculate
{\small
\begin{align*}
&\quad \sigma(j,k)+\sigma(j^\prime,k^\prime)+\sigma(j,2^n-1-k)+\sigma(j^\prime,2^n-1-k^\prime)\\
&= \sum_{m=0}^{n-1} \left(j_m k_m + j_m^\prime k_m^\prime + j_m (1-k_m) + j_m^\prime (1-k_m^\prime)\right)\\
&= \sum_{m=0}^{n-1} \left(j_m + j_m^\prime \right),
\end{align*}
}

\noindent for any $j,j^\prime,k,k^\prime=0,1,\ldots,2^n-1$. 
Therefore, if $j,j^\prime$ satisfy 
{\small
$$
\sum_{m=0}^{n-1} (j_m + j^\prime_m) \mod 2 = 1,
$$
}

\noindent then we have $(-1)^{\sigma(j,k)+\sigma(j^\prime,k^\prime)+\sigma(j,2^n-1-k)+\sigma(j^\prime,2^n-1-k^\prime)}=-1$ for any $k,k^\prime=0,1,\ldots,2^n-1$, and hence $\alpha_{j,j^\prime}=0$. 

Moreover, if we set $\theta_{j,j^\prime} = - v(|j-j^\prime|\Delta x)\Delta t $, then
{\small
\begin{align*}
\theta_{2^n-1-j,2^n-1-j^\prime} &= - v(|2^n-1-j-(2^n-1-j^\prime)|\Delta x)\Delta t \\
&= - v(|j-j^\prime|\Delta x)\Delta t = \theta_{j,j^\prime}, 
\end{align*}
}

\noindent and thus Eq. \eqref{intro:eq-theta2} is a sufficient (but not necessary) condition of the assumption in Theorem 1. 
That is, we have the following corollary. 

\noindent {\textbf{Corollary 1. }}
Let $n\in \mathbb{N} := \{1,2,\ldots\}$. Assume that there exist $2^n$ constants $\Theta_0,\Theta_1,\ldots, \Theta_{2^n-1}$ such that $\theta_{k,k^\prime}$ satisfies
{\small
$$
\theta_{k,k^\prime} = \Theta_{|k-k^\prime|}, \quad k,k^\prime = 0,1,\ldots,2^n-1. 
$$
}

\noindent Then $\alpha_{j,j^\prime}$, defined by Eq. \eqref{app:eq-alpha}, fulfills 
{\small
$$
\alpha_{j,j^\prime} = 0 \quad \text{for any $j,j^\prime$ such that }\quad \sum_{m=0}^{n-1} (j_m + j^\prime_m) \mod 2 = 1. 
$$ 
}

\section{Details on circuit simplification} 
\label{app:subsec-2}

We use a simple example of $n=2$ to illustrate the step-by-step simplification of the circuit 
in Sect.~\ref{sec:3} by using some rewriting rules for the CNOT gates. 

We start from the quantum circuit \cite{ZHL22} and first reduce the $R_z$ gates by Theorem 1, which is shown in Fig.~\ref{pre:fig1}. Note that the number of $R_z$ gates is reduced from $15$ to $7$ (from $2^{2n}-1$ to $2^{2n-1}-1$ in general). 

\begin{figure}
\centering
\resizebox{13.0cm}{!}{
\includegraphics[keepaspectratio]{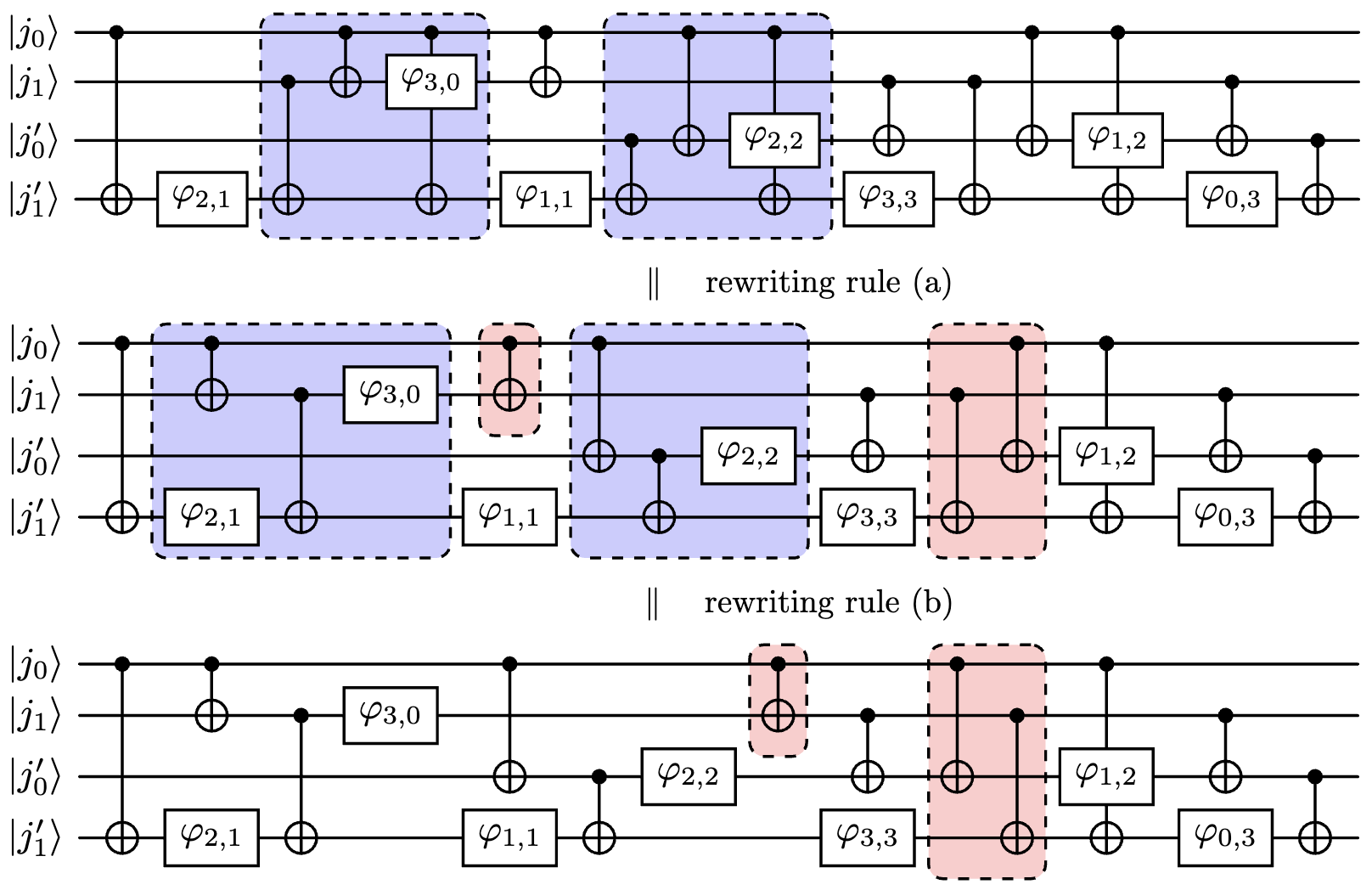}
}%
\caption{(Color online) Reduction of CNOT gates by rewriting rules (a), (b). 
The application of rewriting rule (a)/(b) is shown in blue/red boxes.} 
\label{app:figsim1}
\end{figure}

\begin{figure}
\centering
\resizebox{13.0cm}{!}{
\includegraphics[keepaspectratio]{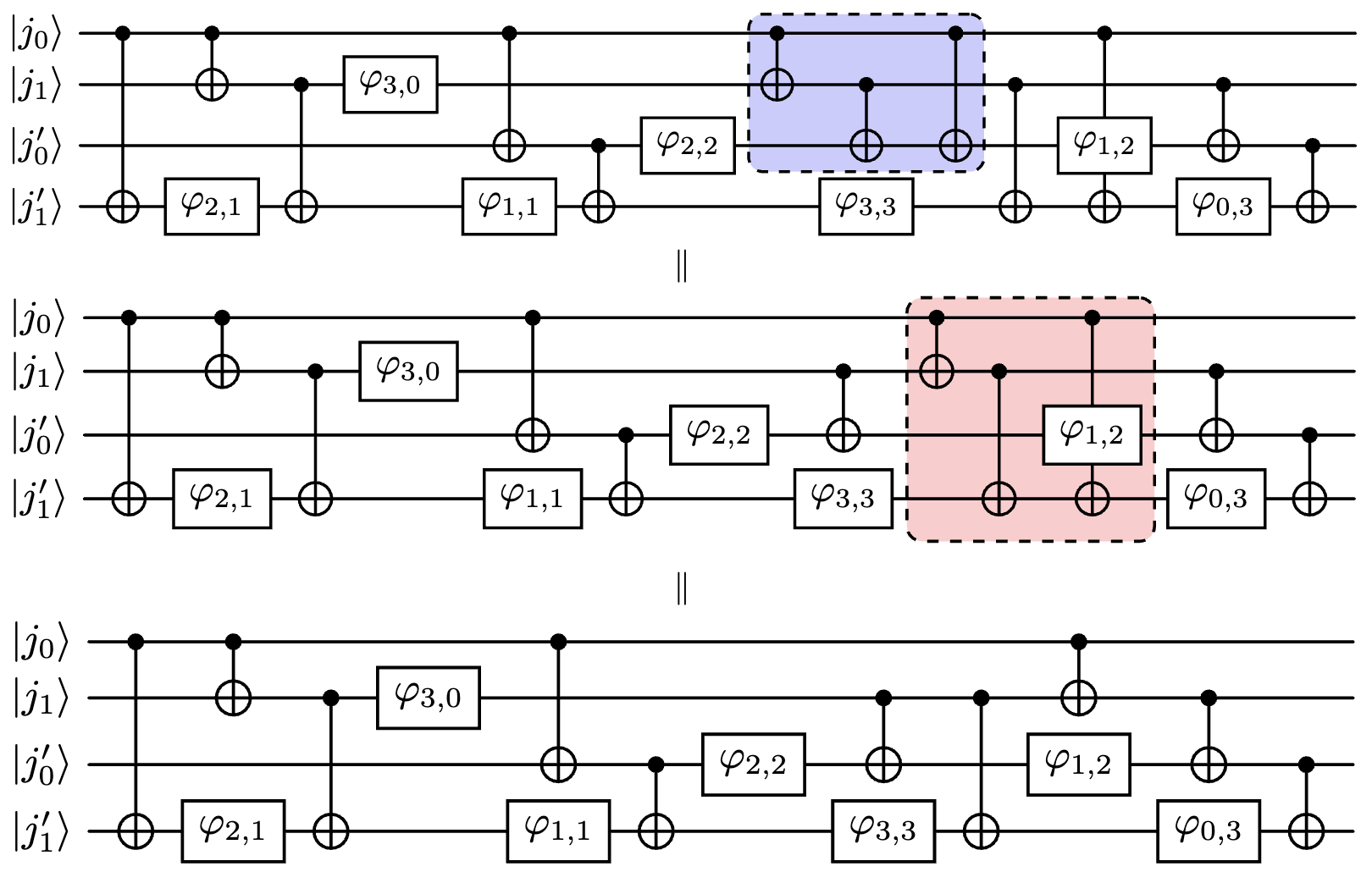}
}%
\caption{(Color online) Reduction of CNOT gates by applying rewriting rule (a) twice. 
The box we apply rewriting rule (a) for the first time/the second time is highlighted in blue/red.} 
\label{app:figsim2}
\end{figure}

Next, we apply sequentially (i) the rewriting rule (a), (ii) the rewriting rule (b), (iii) the rewriting rule (a) in Fig.~\ref{pre:fig2} to reduce the CNOT gates (see Figs. \ref{app:figsim1}--\ref{app:figsim2}). 
Note that the number of CNOT gates is reduced from $14$ to $10$ (from $2^{2n}-2$ to $2^{2n-1}+2n-2$ in general). 

\begin{figure}
\centering
\resizebox{13.0cm}{!}{
\includegraphics[keepaspectratio]{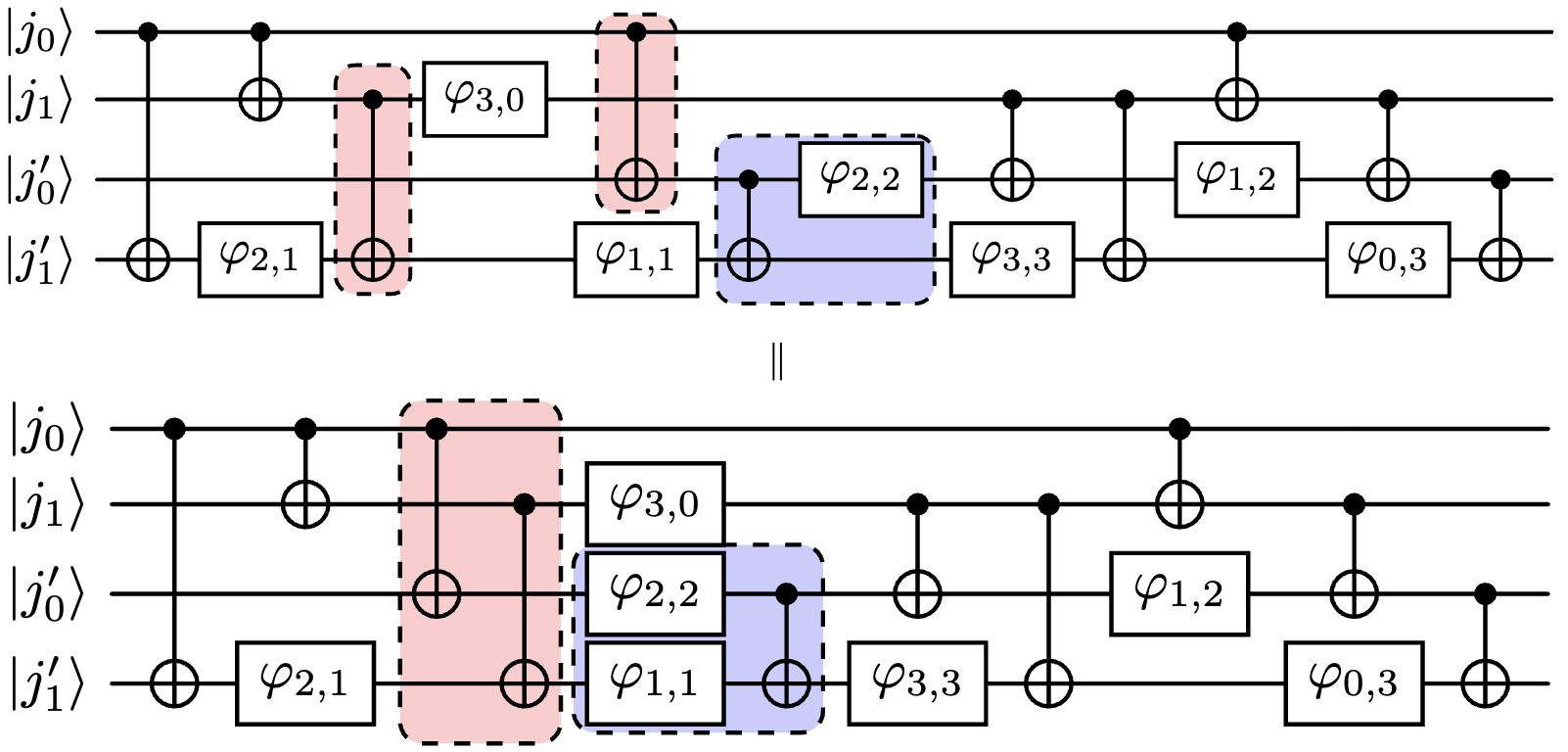}
}%
\caption{(Color online) Reduction of circuit depth by applying rewriting rules (b), (c). 
The application of rewriting rule (b)/(c) is shown in blue/red boxes. }
\label{app:figsim3}
\end{figure}

Finally, we suitably employ the rewriting rules (b) and (c) to reduce further the circuit depth, which is demonstrated in Fig.~\ref{app:figsim3}. 
Note that the circuit depth is reduced from $16$ to $10$. 


\section{Validation of Algorithm 1}
\label{app:subsec-3}

For simplicity, we illustrate this by an example of $n=2$. 
First, we recall that the ordering given by Algorithm 1 is as follows:
{\small
\begin{align*}
&\mathrm{CNOT}(3,0), \mathrm{CNOT}(2,0), \mathrm{CNOT}(1,0), \mathrm{F}(3,1,\varphi_{1001}), \mathrm{F}(3,2,\varphi_{0101}), \\
&\mathrm{F}(2,1,\varphi_{1010}), 
\mathrm{F}(3,1,\varphi_{1111}), \mathrm{F}(1,0,\varphi_{1100}), \mathrm{F}(2,1,\varphi_{0110}), 
\mathrm{F}(3,2,\varphi_{0011}). 
\end{align*}
}

\noindent Starting from the quantum state $\ket{q}=\ket{q_0}\otimes \ket{q_1}\otimes \ket{q_2}\otimes \ket{q_3}$, we apply the CNOT gates in Step 0:
{\small
$$
\ket{q} \xrightarrow{\text{Step 0}} \ket{q_0} \otimes \ket{q_1\oplus q_0} \otimes \ket{q_2\oplus q_0} \otimes \ket{q_3\oplus q_0}.
$$
}

\noindent In Step 1, we apply the block $\mathrm{B}_2 = \mathrm{F}(3,2,\varphi_{0101})\mathrm{F}(3,1,\varphi_{1001})$ (we use the notation for operators, which means we first apply $\mathrm{F}(3,1,\varphi_{1001})$) to obtain
{\small
\begin{align*}
&\xrightarrow{\mathrm{F}(3,1,\varphi_{1001})} 
\mathrm{exp}\left(-\mathrm{i}\varphi_{1001}(-1)^{q_3\oplus q_0}/2\right)\ket{q_0} \otimes \ket{q_1\oplus q_0} \otimes \ket{q_2\oplus q_0} \\
&\hspace{1.2cm} \otimes \ket{q_3\oplus q_1}\\
&\xrightarrow{\mathrm{F}(3,2,\varphi_{0101})} 
\mathrm{exp}\left(-\mathrm{i}(\varphi_{1001}(-1)^{q_3\oplus q_0}+\varphi_{0101}(-1)^{q_3\oplus q_1})/2\right) \\
&\hspace{1.2cm} \ket{q_0} \otimes \ket{q_1\oplus q_0} \otimes \ket{q_2\oplus q_0} \otimes \ket{q_3\oplus q_2\oplus q_1\oplus q_0}.
\end{align*}
}

\noindent In Step 2, we apply the block $\mathrm{B}_1 = \mathrm{F}(3,1,\varphi_{1111})\mathrm{F}(2,1,\varphi_{1010})$ to obtain
{\small
\begin{align*}
&\xrightarrow{\mathrm{F}(2,1,\varphi_{1010})} 
\mathrm{exp}\left(-\mathrm{i}(\varphi_{1001}(-1)^{q_3\oplus q_0}\!+\!\varphi_{0101}(-1)^{q_3\oplus q_1}\!+\!\varphi_{1010}(-1)^{q_2\oplus q_0})/2\right)\\
&\hspace{1.2cm}\ket{q_0} \otimes \ket{q_1\oplus q_0} \otimes \ket{q_2\oplus q_1} \otimes \ket{q_3\oplus q_2\oplus q_1\oplus q_0}\\
&\xrightarrow{\mathrm{F}(3,1,\varphi_{1111})} 
\mathrm{exp}\left(-\mathrm{i}(\varphi_{1001}(-1)^{q_3\oplus q_0}\!+\!\varphi_{0101}(-1)^{q_3\oplus q_1}\!+\!\varphi_{1010}(-1)^{q_2\oplus q_0})/2\right)\\
&\hspace{1.2cm} \times \mathrm{exp}\left(-\mathrm{i}\varphi_{1111}(-1)^{q_3\oplus q_2\oplus q_1\oplus q_0}/2\right) \\
&\hspace{1.2cm} \ket{q_0} \otimes \ket{q_1\oplus q_0} \otimes \ket{q_2\oplus q_1} \otimes \ket{q_3\oplus q_2}.
\end{align*}
}

\noindent In Step 3, we apply $\mathrm{F}(3,2,\varphi_{0011})\mathrm{F}(2,1,\varphi_{0110})\mathrm{F}(1,0,\varphi_{1100})$ and finally obtain
{\small
\begin{align*}
&\xrightarrow{\mathrm{F}(1,0,\varphi_{1100})} 
\mathrm{exp}\left(-\mathrm{i}(\varphi_{1001}(-1)^{q_3\oplus q_0}\!+\!\varphi_{0101}(-1)^{q_3\oplus q_1}\!+\!\varphi_{1010}(-1)^{q_2\oplus q_0})/2\right)\\
&\hspace{1.2cm} \times \mathrm{exp}\left(-\mathrm{i}(\varphi_{1111}(-1)^{q_3\oplus q_2\oplus q_1\oplus q_0}\!+\!\varphi_{1100}(-1)^{q_1\oplus q_0})/2\right)\\
&\hspace{1.2cm} \ket{q_0} \otimes \ket{q_1} \otimes \ket{q_2\oplus q_1} \otimes \ket{q_3\oplus q_2}\\
&\xrightarrow{\mathrm{F}(2,1,\varphi_{0110})} 
\mathrm{exp}\left(-\mathrm{i}(\varphi_{1001}(-1)^{q_3\oplus q_0}\!+\!\varphi_{0101}(-1)^{q_3\oplus q_1}\!+\!\varphi_{1010}(-1)^{q_2\oplus q_0})/2\right)\\
&\hspace{1.2cm} \times \mathrm{exp}\left(-\mathrm{i}(\varphi_{1111}(-1)^{q_3\oplus q_2\oplus q_1\oplus q_0}\!+\!\varphi_{1100}(-1)^{q_1\oplus q_0})/2\right) \\
&\hspace{1.2cm} \times \mathrm{exp}\left(-\mathrm{i}\varphi_{0110}(-1)^{q_2\oplus q_1}/2\right) \ket{q_0} \otimes \ket{q_1} \otimes \ket{q_2} \otimes \ket{q_3\oplus q_2}\\
&\xrightarrow{\mathrm{F}(3,2,\varphi_{0011})} 
\mathrm{exp}\left(-\mathrm{i}(\varphi_{1001}(-1)^{q_3\oplus q_0}\!+\!\varphi_{0101}(-1)^{q_3\oplus q_1}\!+\!\varphi_{1010}(-1)^{q_2\oplus q_0})/2\right)\\
&\hspace{1.2cm} \times \mathrm{exp}\left(-\mathrm{i}(\varphi_{1111}(-1)^{q_3\oplus q_2\oplus q_1\oplus q_0}\!+\!\varphi_{1100}(-1)^{q_1\oplus q_0})/2\right) \\
&\hspace{1.2cm} \times \mathrm{exp}\left(-\mathrm{i}(\varphi_{0110}(-1)^{q_2\oplus q_1}\!+\!\varphi_{0011}(-1)^{q_3\oplus q_2})/2\right) \\
&\hspace{1.2cm} \ket{q_0} \otimes \ket{q_1} \otimes \ket{q_2} \otimes \ket{q_3}.
\end{align*}
}

By Theorem 1, we find that half of the rotation angles $\varphi_{j,j^\prime}$ vanish, and then the terms in the exponential can be written by a summation over $j,j^{\prime}$ from $0$ to $3$ as follows:  
{\small
\begin{align*}
&\ket{q} \longrightarrow e^{\mathrm{i}\varphi_{0,0}/2}\mathrm{exp}\left(-\mathrm{i}\left(\sum_{j,j^\prime=0}^3 \varphi_{j,j^\prime}(-1)^{\sigma(q^{(1)},j)+\sigma(q^{(2)},j^\prime)}/2\right)\right)\ket{q}\\
&= e^{\mathrm{i}\varphi_{0,0}/2} \mathrm{exp}\left(\mathrm{i}\left(\frac{1}{2^{4}}\sum_{j,j^\prime=0}^3 \sum_{k,k^\prime=0}^3 (-1)^{\sigma(q^{(1)},j)+\sigma(q^{(2)},j^\prime)+\sigma(j,k)+\sigma(j^\prime,k^\prime)}\theta_{k,k^\prime}\right)\right)\ket{q}.
\end{align*}
}

\noindent Here, $\ket{q}_{2n} = \ket{q^{(1)}}_n \otimes \ket{q^{(2)}}_n$. If we write two registers together, then we have 
{\small
\begin{align*}
\ket{q} \longrightarrow e^{\mathrm{i}\varphi_{0}/2} \mathrm{exp}\left(\mathrm{i}\left(\frac{1}{2^{4}}\sum_{\tilde j=0}^{15} \sum_{\tilde k=0}^{15} (-1)^{\sigma(q,\tilde j)+\sigma(\tilde j,\tilde k)}\theta_{\tilde k}\right)\right)\ket{q}, 
\end{align*}
}

\noindent where $\tilde j = 2^{2} j + j^\prime$, $j,j^\prime=0,1,2,3$. 
Similarly, for general $n\ge 2$ we assert that
{\small
\begin{align*}
\ket{q} \longrightarrow e^{\mathrm{i}\varphi_{0}/2} \mathrm{exp}\left(\mathrm{i}\left(\frac{1}{2^{2n}}\sum_{\tilde j=0}^{2^{2n}-1} \sum_{\tilde k=0}^{2^{2n}-1} (-1)^{\sigma(q,\tilde j)+\sigma(\tilde j,\tilde k)}\theta_{\tilde k}\right)\right)\ket{q}. 
\end{align*}
}

Again, we use the matrix $\mathbf{H}\in \mathbb{R}^{2^{2n}\times 2^{2n}}$ of $2n$-qubit Hadamard transform and note that $\bm{\alpha} = \mathbf{H} \bm{\theta}$ can be written in components:
{\small
$$
\alpha_{\tilde j} = \frac{1}{\sqrt{2^{2n}}}\sum_{\tilde k=0}^{2^{2n}-1}(-1)^{\sigma(\tilde j,\tilde k)\theta_{\tilde k}}, \quad \tilde j,\tilde k=0,1,\ldots,2^{2n}-1. 
$$
}

\noindent Then we have
{\small
$$
\frac{1}{2^{2n}}\sum_{\tilde j=0}^{2^{2n}-1} \sum_{\tilde k=0}^{2^{2n}-1} (-1)^{\sigma(q,\tilde j)+\sigma(\tilde j,\tilde k)}\theta_{\tilde k}
= \frac{1}{\sqrt{2^{2n}}}\sum_{\tilde j=0}^{2^{2n}-1} (-1)^{\sigma(q,\tilde j)} \alpha_{\tilde j} = \theta_{q},
$$
}

\noindent where we used $\bm{\theta} = \mathbf{H}^2 \bm{\theta} = \mathbf{H} \bm{\alpha}$. 
Therefore, by Algorithm 1, we obtain
{\small
\begin{align*}
\ket{q} \longrightarrow e^{\mathrm{i}\varphi_{0}/2} e^{\mathrm{i}\theta_q} \ket{q}, \quad q=0,1,\ldots,2^{2n}-1, 
\end{align*}
}

\noindent which completes the implementation of the desired diagonal matrix $\mathrm{D}(\bm{\theta})$ up to a global phase $e^{\mathrm{i}\varphi_{0}/2}$. 
Therefore, we have the following theorem:

\noindent {\textbf{Theorem 2. }}
Let $\alpha_{j,j^\prime}$ be defined by Eq. \eqref{app:eq-alpha}. 
If $\alpha_{j,j^\prime}$ satisfies
{\small
$$
\alpha_{j,j^\prime} = 0 \quad \text{for any $j,j^\prime$ such that }\quad \sum_{m=0}^{n-1} (j_m + j^\prime_m) \mod 2 = 1, 
$$ 
}

\noindent then Algorithm 1 gives an implementation of 
{
$$
e^{\mathrm{i}\varphi_0/2}\mathrm{D}(\bm{\theta}),
$$
}

\noindent where $\varphi_0 = -2\alpha_{0,0}/\sqrt{2^{2n}}$. 

\begin{figure}
\centering
\resizebox{13.0cm}{!}{
\includegraphics[keepaspectratio]{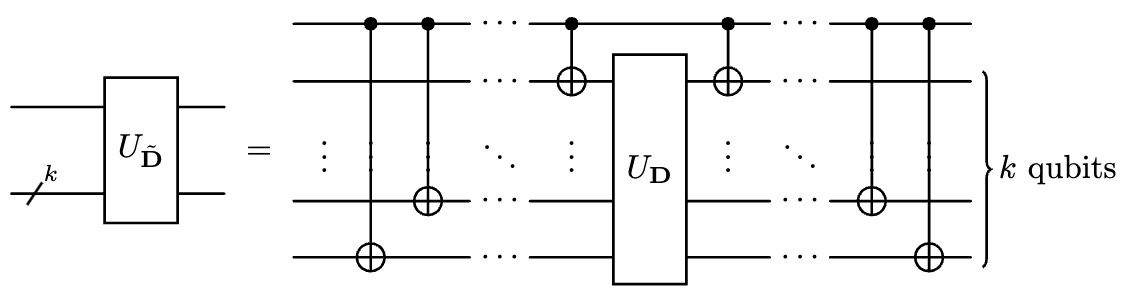}
}%
\caption{Another implementation of the diagonal unitary matrix with reflection symmetry. }
\label{app:fig0}
\end{figure}

We remark that there is another way to implement the diagonal unitary matrices with reflection symmetry fulfilling Eq. \eqref{intro:eq-theta}. 
Let $\mathbf{D}$ be a given $2^k\times 2^k$ diagonal unitary matrix and $\mathbf{D}_{\mathrm{ref}}$ be its reflection (the matrix that the diagonal entries are in reverse order). We introduce an augmented matrix defined by
$$
\mathbf{\tilde D} = 
\begin{pmatrix}
\mathbf{D} & 0 \\
0 & \mathbf{D_{\mathrm{ref}}}
\end{pmatrix}.
$$
Moreover, let $U_{\mathbf{D}}$ and $U_{\mathbf{\tilde D}}$ be the operators implementing the diagonal unitary matrices $\mathbf{D}$ and $\mathbf{\tilde D}$, respectively. Then, we have the equivalence of the circuit in Fig.~\ref{app:fig0}. 

In fact, if we denote the block with $k$ CNOT gates in Fig.~\ref{app:fig0} by $\mathcal{C}$ and let 
{\small
$$
\mathbf{D} = \sum_{j=0}^{2^k-1} e^{\mathrm{i} a_j}\ket{j}\bra{j},
$$
}

\noindent where $a_j\in \mathbb{R}$ for $j=0,1,\ldots,2^k-1$, $\ket{\hat q}_k = \ket{q_1}\otimes \cdots \otimes \ket{q_k}$, then we have
{\small
\begin{align*}
&\quad \mathcal{C}(I \otimes U_{\mathbf{D}}) \mathcal{C} (\ket{q_0} \otimes \ket{\hat q}_k) \\
&= \mathcal{C}(I \otimes U_{\mathbf{D}}) (\ket{q_0} \otimes \ket{q_1\oplus q_0} \otimes \cdots \otimes \ket{q_k\oplus q_0}) \\
&= \mathcal{C} (\ket{q_0} \otimes e^{\mathrm{i} a_{[q_1\oplus q_0 \cdots q_k\oplus q_0]}}\ket{q_1\oplus q_0} \otimes \cdots \otimes \ket{q_k\oplus q_0}) \\
&= e^{\mathrm{i} a_{[q_1\oplus q_0 \cdots q_k\oplus q_0]}} \ket{q_0} \otimes \ket{\hat q}_k, 
\end{align*}
}

\noindent and hence
{\small
\begin{align*}
\mathcal{C}(I \otimes U_{\mathbf{D}}) \mathcal{C} 
&= e^{\mathrm{i} a_{[q_1 \cdots q_k]}} \ket{0}\bra{0} \otimes \ket{\hat q}\bra{\hat q}_k + e^{\mathrm{i} a_{[q_1\oplus 1 \cdots q_k\oplus 1]}} \ket{1}\bra{1} \otimes \ket{\hat q}\bra{\hat q}_k \\
&= e^{\mathrm{i} a_{\hat q}} \ket{0}\bra{0} \otimes \ket{\hat q}\bra{\hat q}_k + e^{\mathrm{i} a_{2^k-1-\hat q}} \ket{1}\bra{1} \otimes \ket{\hat q}\bra{\hat q}_k = U_{\mathbf{\tilde D}}.
\end{align*}
}

\noindent By applying the result in Ref. \cite{ZHL22} for the circuit of $U_{\mathbf{D}}$ ($k=2n-1$), we manage to implement the diagonal unitary matrices with reflection symmetry by a circuit with $2^{2n-1}-2+2(2n-1)=2^{2n-1}+4n-4$ CNOT gates, which has $2n-2$ more CNOT gates than that by Algorithm 1. 

In the end, we give a remark on the idea of the construction of a diagonal matrix. For a general $2^m\times 2^m$ diagonal matrix, one can find a direct implementation by using $2^m$ multifold $z$ rotation gates, which is the base of the $\{\mathrm{CNOT}, R_z\}$ implementation methods (see e.g.,\! Refs. \cite{W.14, ZHL22}). At the same time, one can calculate the rotation angle $\varphi_j$ for each rotation gate. Then, the key point is to implement suitably the sub-blocks: 
{\small
$$
\ket{q} \longrightarrow e^{\mathrm{i}(-1)^{\sigma(j,q)}\varphi_j}\ket{\tilde q},
$$
}

\noindent for $j=0,1,\ldots,2^m-1$. Here, $\ket{\tilde q}$ can be different from $\ket{q}$, but one needs to retrieve the initial state $\ket{q}$ at last by applying suitable CNOT gates. The existing papers, including this paper, follow this idea and consider optimal implementation in circuit depth or gate count. 
In this paper, we manage to reduce both the depth and entanglement gate count thanks to the symmetric property, which reduces the rotation gates (Theorem 1). 
As a result, we can skip several sub-blocks above to obtain a more efficient implementation by finding a way to realize the rest sub-blocks. 


\section{Quantum circuit by using a quantum comparator and modular subtractor}
\label{app:subsec-4}

In the main contents, we focus on the direct implementation (without any ancillas) of the desired diagonal unitary matrix $\mathrm{D}(\bm{\theta})$ satisfying Eq. \eqref{intro:eq-theta}. 
As we mentioned in Sect.~\ref{sec:6}, there is another method for the special case of Eq. \eqref{intro:eq-theta2} by a quantum comparator and a modular subtractor.  
Since such an example has potential application in the problem of the real-time evolution of the first quantized Hamiltonian for interacting particles, we provide a detailed discussion of the quantum circuit here.
Among the existing papers on quantum comparators and modular subtractors \cite{BKS.18, D20,Li.20, OOCFG21, OR07, S20, TRF10, TK06, YWW.23}, we choose the ones in Ref. \cite{Li.20} to explain our idea as they provide better circuits than the other existing results. 

\begin{figure}
\centering
\resizebox{13.0cm}{!}{
\includegraphics[keepaspectratio]{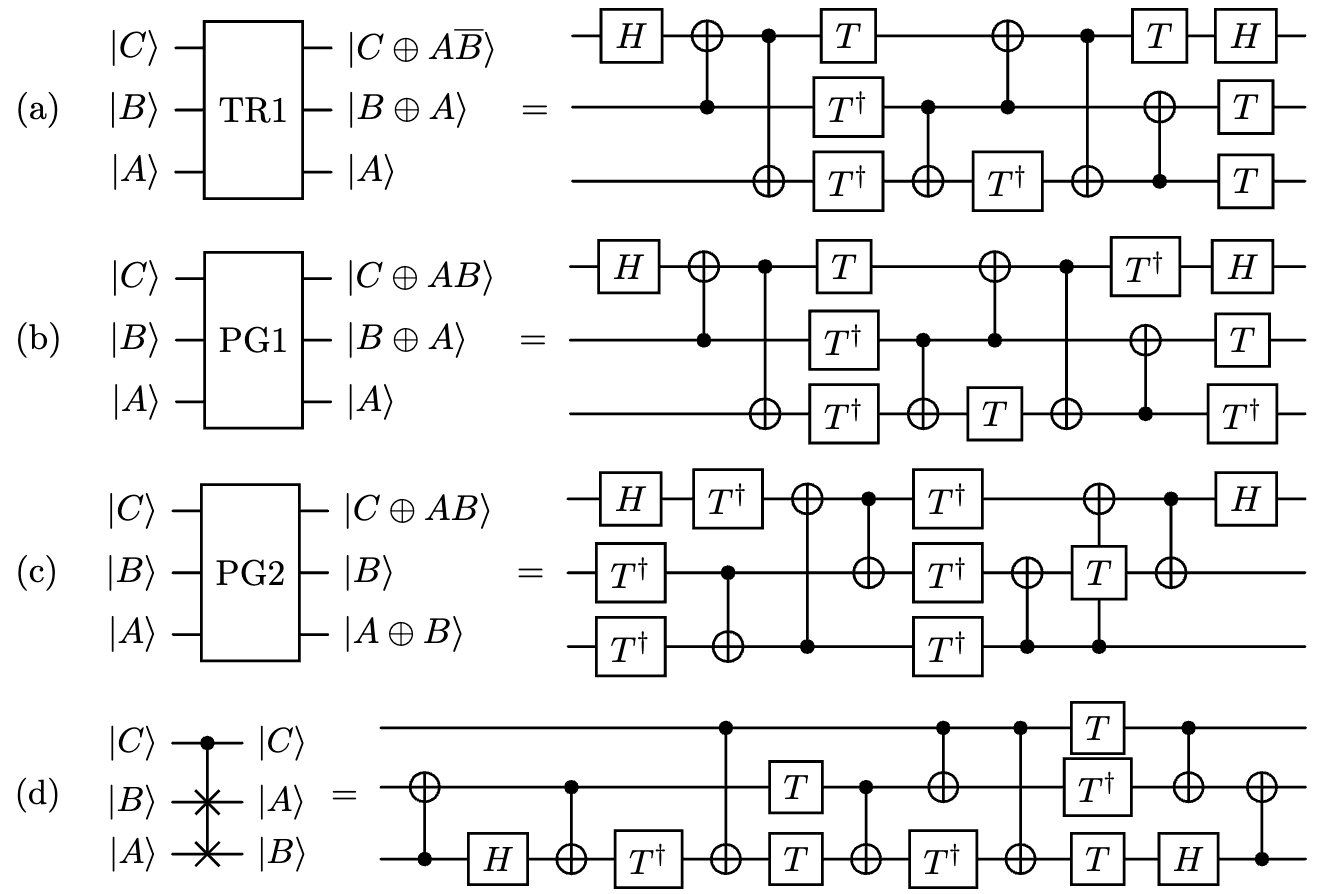}
}%
\caption{TR gate, Peres gates and their implementations with $6$ CNOT gates and depth $9$ \cite{Li.20}. (a) TR gate $\mathrm{TR1}$, (b) Peres gate $\mathrm{PG1}$, (c) Peres gate $\mathrm{PG2}$, (d) Fredkin gate.}
\label{app:fig1}
\end{figure}

First, we introduce the gates $\mathrm{TR1}$, $\mathrm{PG1}$ and $\mathrm{PG2}$ \cite{Li.20} (see Fig.~\ref{app:fig1}). Note that each gate has $6$ CNOT gates with a circuit depth of $9$. 

\begin{figure*}[htb]
\centering
\resizebox{13cm}{!}{%
\includegraphics[keepaspectratio]{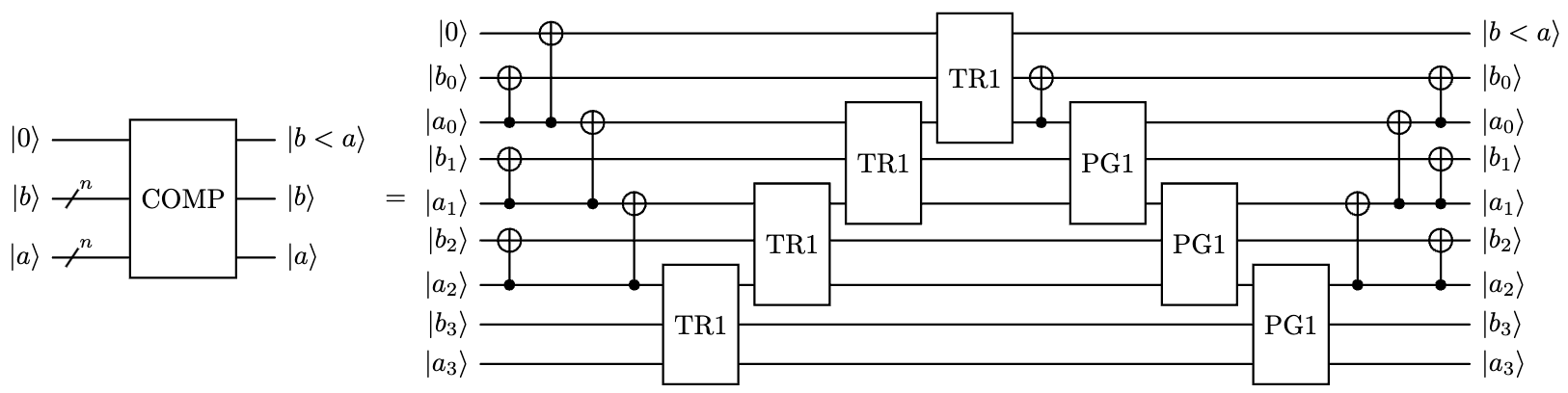}
}%
\caption{Quantum comparator (Fig.~11\cite{Li.20}) for $\ket{a}=\ket{a_0 a_1 a_2 a_3}$ and $\ket{b}=\ket{b_0 b_1 b_2 b_3}$ in the case of $n=4$.}
\label{app:fig-comp}
\end{figure*}

\begin{figure*}[htb]
\centering
\resizebox{13cm}{!}{%
\includegraphics[keepaspectratio]{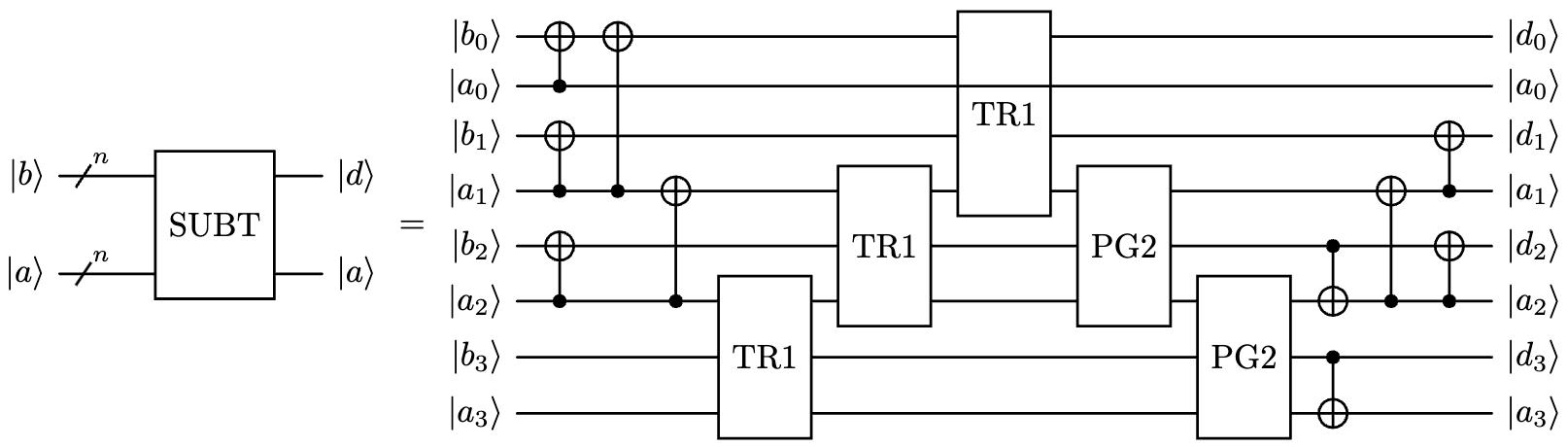}
}%
\caption{Quantum modular subtractor (Fig.~10\cite{Li.20}) for  $\ket{d}=\ket{(b-a)}$ in the case of $n=4$.}
\label{app:fig-subt}
\end{figure*}

Next, the quantum comparator and the modular subtractor are illustrated by Figs. \ref{app:fig-comp} and \ref{app:fig-subt}, respectively. 
For simplicity, we provide the circuits for only $n=4$, but it is readily to check that in the case of general $n$, the quantum comparator (Fig.~11\cite{Li.20}) has $4n-4$ CNOT gates, $n$ $\mathrm{TR1}$ gates and $n-1$ $\mathrm{PG1}$ gates, and hence has totally $16n-10$ for $n\ge 2$ CNOT gates. 
On the other hand, the quantum modular subtractor (Fig.~10\cite{Li.20}) has $5n-10$ CNOT gates for $n\ge 3$ ($1$ CNOT gate for $n=2$), $n-1$ $\mathrm{TR1}$ gates and $n-2$ $\mathrm{PG2}$ gates, and hence has totally $17n-28$ CNOT gates for $n\ge 3$ ($7$ CNOT gates for $n=2$). 
Moreover, we can estimate the upper bound of circuit depth for the comparator: 
{\small
$$
d_{\mathrm{COMP}} \le 2n+9(2n-1) = 20n-9,\quad n\ge 2,
$$
}

\noindent and the upper bound for the modular subtractor: 
{\small
$$
d_{\mathrm{SUBT}} 
\left\{
\begin{aligned}
& \le 2(n-1)+9(2n-3) = 20n-29,\quad n\ge 3,\\
& = 10,\quad n=2.
\end{aligned}
\right.
$$
}
Finally, in terms of the circuit in Ref. \cite{ZHL22}, denoted by $U_{\mathrm{D}}$, the Fredkin gates and the above comparator as well as modular subtractor, we can implement the desired diagonal unitary matrix $\mathrm{D}(\bm{\theta})$ satisfying Eq. \eqref{intro:eq-theta2} by the circuit in Fig.~\ref{dis:fig1}. 
The circuit involves the uncomputation part, which enables one to discard or reuse the ancilla safely by disentangling it from the other qubits.


\end{document}